\documentclass[useAMS,usenatbib]{mnras}

\usepackage{graphics}
\usepackage{amssymb}
\usepackage{amsfonts}
\usepackage{wasysym}
\usepackage{epsfig}
\usepackage{breakurl}

\usepackage{longtable}
\usepackage{times}
\usepackage{mathptmx}
\usepackage[usenames,dvipsnames]{color}
\usepackage{url}

\title[Radio emission from ordered magnetospheres]
{A scaling relationship for non-thermal radio emission from ordered magnetospheres: from the top of the Main Sequence to planets}
\author[P. Leto et al.]
{P. Leto,$^{1}$ \thanks{E-mail: paolo.leto@inaf.it}
C. Trigilio,$^{1}$
J. Krti\v{c}ka,$^{2}$ 
L. Fossati,$^{3}$ 
R. Ignace,$^{4}$ 
M. E. Shultz,$^{5}$
C. S. Buemi,$^{1}$ 
\newauthor 
L. Cerrigone,$^{6}$ 
G. Umana,$^{1}$
A. Ingallinera,$^{1}$
C. Bordiu,$^{1,7}$ 
I. Pillitteri,$^{8}$ 
F. Bufano,$^{1}$
\newauthor 
L. M. Oskinova,$^{9,10}$ 
C. Agliozzo,$^{11}$ 
F. Cavallaro,$^{12,1}$ 
S. Riggi,$^{1}$
S. Loru,$^{1}$
H. Todt,$^{9}$ 
\newauthor 
M. Giarrusso,$^{13}$ 
N. M. Phillips,$^{11}$ 
J. Robrade,$^{14}$ 
F. Leone$^{15,1}$
\\
$^{1}$INAF -- Osservatorio Astrofisico di Catania, Via S. Sofia 78, I-95123 Catania, Italy\\ 
$^2$Department of Theoretical Physics and Astrophysics, Masaryk University, Kotl\'{a}\v{r}sk\'{a} 2, CZ-611 37 Brno, Czech Republic\\ 
$^3$Space Research Institute, Austrian Academy of Sciences, Schmiedlstrasse 6, A-8042 Graz, Austria\\ 
$^4$Department of Physics \& Astronomy, East Tennessee State University, Johnson City, TN 37614, USA\\ 
$^5$Department of Physics and Astronomy, and SARA, University of Delaware, 217 Sharp Lab, Newark, DE 19716, USA\\
$^6$Joint ALMA Observatory, Alonso de C\'{o}rdova 3107, Vitacura, 8320000, Santiago, Chile\\  
$^{7}$Centro de Astrobiolog\'{i}a (INTA-CSIC), Ctra. M-108, km. 4, Torrej\'{o}n de Ardoz, E-28850 Madrid, Spain\\
$^{8}$INAF -- Osservatorio Astronomico di Palermo, Piazza del Parlamento 1, I-90134 Palermo, Italy\\ 
$^9$Institute for Physics and Astronomy, University Potsdam, D-14476 Potsdam, Germany\\ 
$^{10}$Kazan Federal University, Kremlevskaya Str 18, 42008, Kazan, Russia\\ 
$^{11}$European Southern Observatory, Karl-Schwarzschild-Strasse 2, D-85748 Garching bei M\"{u}nchen, Germany\\ 
$^{12}$The Inter-University Institute for Data Intensive Astronomy (IDIA), Department of Astronomy, University of Cape Town, Rondebosch, 7701, South Africa\\ 
$^{13}$INFN, Laboratori Nazionali del Sud, Via S. Sofia 62, I-95123 Catania, Italy\\ 
$^{14}$Hamburger Sternwarte, University of Hamburg, Gojenbergsweg 112, D-21029 Hamburg, Germany\\ 
$^{15}$Dipartimento di Fisica e Astronomia, Sezione Astrofisica, Universit\'{a} di Catania, Via S. Sofia 78, I-95123 Catania, Italy\\ 
}
\begin{document}

\date{}

\pagerange{\pageref{firstpage}--\pageref{lastpage}} \pubyear{}

\maketitle

\label{firstpage}

\begin{abstract}
In this paper, we present the analysis of incoherent non-thermal radio emission from a sample of hot magnetic stars, ranging from early-B to early-A spectral type. Spanning a wide range of stellar parameters and wind properties, these stars display a commonality in their radio emission which presents new challenges to the wind scenario as originally conceived. It was thought that relativistic electrons, responsible for the radio emission, originate in current sheets formed where the wind opens the magnetic field lines. However, the true mass-loss rates from the cooler stars are too small to explain the observed non-thermal broadband radio spectra. Instead, we suggest the existence of a radiation belt located inside the inner-magnetosphere, similar to that of Jupiter. Such a structure explains the overall indifference of the broadband radio emissions on wind mass-loss rates. Further, correlating the radio luminosities from a larger sample of magnetic stars with their stellar parameters, the combined roles of rotation and magnetic properties have been empirically determined. Finally, our sample of early-type magnetic stars suggests a scaling relationship between the non-thermal radio luminosity and the electric voltage induced by the magnetosphere's co-rotation, which appears to hold for a broader range of stellar types with dipole-dominated magnetospheres (like the cases of the planet Jupiter and the ultra-cool dwarf stars and brown dwarfs). We conclude that well-ordered and stable rotating magnetospheres share a common physical mechanism for supporting the generation of non-thermal electrons.
\end{abstract}


%
%
%
\section{Introduction}

The magnetic fields of magnetic main-sequence B/A type stars have, 
on average, a strength of a few kG \citep{sikora_etal19b,shultz_etal19_490} with stable field topologies
\citep{shultz_etal18}.
The magnetic field topology of the majority of them can be described in terms 
of the oblique rotator model (ORM), predominantly involving a 
centered dipole field that is tilted to the stellar rotation axis 
\citep{babcock49,stibbs50}. The presence of a large-scale magnetic field 
induces surface chemical anisotropies on the stellar surface, making 
these stars variable as a consequence of stellar rotation
\citep{krticka_etal07}. 
In general early-type magnetic stars are recognized based on 
their chemical peculiarities, and are classified as Ap/Bp stars \citep{preston74}.

The MiMeS (Magnetism in Massive Stars; \citealp{wade_etal16}) 
and BOB (B  fields  in  OB stars; \citealp{morel_etal15})
spectropolarimetric surveys concluded that magnetism
characterizes about $6$--8\% of the stars across the upper
main-sequence (MS) \citep{fossati_etal15,wade_etal16,grunhut_etal17,sikora_etal19a},
with the incidence rate almost constant for stars ranging from the O to the Ap stars.     
The magnetism
almost certainly has a fossil origin \citep{braithwaite_spruit04,duez_mathis10,Neiner2015} rather than being created by
a magnetic dynamo as in late-type stars; this is supported by the
magnetic field decay during the stellar life 
\citep{landstreet_etal07,
landstreet_etal08,
fossati_etal16,
sikora_etal19b,
petit_etal19,
shultz_etal19_490}.
Further, stellar mergers have been proposed as 
origin of the strong magnetic fields observed at the top of the MS
\citep{bogomazov_tutukov09,schneider_etal16,schneider_etal19,keszthelyi_etal21}.

The early-type magnetic stars are sufficiently hot 
to radiatively drive
stellar wind, which in the presence of their
large-scale magnetic fields may be strongly aspherical
\citep{babel_montmerle97,ud-doula_owocki02}. 
The wind
plasma accumulates at low magnetic latitudes (inner magnetosphere),
whereas it freely propagates along directions aligned with the magnetic
poles \citep{shore87,shore_brown90,leone93}.

This concept has led to the magnetically
confined wind shock (MCWS) model \citep{babel_montmerle97}.  In the
presence of large-scale magnetic fields, the wind cannot freely 
propagate spherically, and the ionized wind flow is channeled by
the magnetic field lines. 
At the magnetic equator, wind streams
arising from the opposite magnetic hemispheres collide and {shock},
producing X-rays \citep{oskinova_etal11,naze_etal14,robrade16}.
Detailed modeling of the magnetospheres of hot magnetic stars
has progressively developed
\citep*{ud-doula_owocki02,ud-doula_etal06,ud-doula_etal08,ud-doula_etal13,
ud-doula_etal14,owocki_etal16,Daley-Yates_etal19,munoz_etal20}.

{Rapidly rotating magnetic stars require the presence of strong magnetic fields to enable formation of centrifugal magnetospheres}
\citep{maheswaran_cassinelli09,petit_etal13,shultz_etal19_490}, where the centrifugal force
is strong enough to balance the gravitational infall of the circumstellar
ionized material.
The magnetospheric ionized matter 
is forced to rigidly
co-rotate with the stars, producing a rigidly rotating magnetosphere
(RRM), which is characterized by typical signatures recognized in
the H$\alpha$ line profile and photometric variability
\citep*{townsend_owocki05,townsend08}.

\begin{table*}
\caption[ ]{Stars of the sample and corresponding stellar parameters.}
\label{tab_stars}
\footnotesize
\begin{tabular}{@{}r@{~~~~} l@{~~} c@{~~} c@{~~} c@{~~} l@{~~~} l @{~~~}l@{~~~} l@{~~~} l@{~~~~~} c@{}}
\hline
       &                      &                                &   &$D$                     &$T_{\mathrm{eff}}$ &$R_{\ast}$      &$P_{\mathrm{rot}}^{\dag}$  &$B_{\mathrm p}$  &Frac. MS age &$L_{\nu,\mathrm{rad}}$       \\
ID        &HD                &Alt. name              & Sp. type                 &(pc)                      &(kK)                       &(R$_{\odot}$)   &(d)                            &(kG)                      &                      &(erg s$^{-1}$ Hz$^{-1}$)     \\
\hline                           
1       &12447             &$\alpha$\,Psc\,A      &A2 SiSrCr   &$~~50(2)$               &$10.2 (0.5)$   &$2.6 (0.2 )^{(3)}$  &1.4907(8)$^{(17)}$           &$1.19\pm0.2^{(33)}$     &$0.58 (0.2 )^{(36)}$       & $1.8(0.6)\times 10^{15}$    \\
2       &19832             &56\,Ari               &B8 Si       &$~~124(3)~~$            &$12.5 (0.5)$   &$2.3 (0.3)^{(4)}$  &0.72795(1)$^{(18)}$          & $2.7^{+0.6}_{-0.3}$$^{(4)}$     &$0.51 (0.26 )^{(36)}$        & $~~~~~~9(2)\times 10^{15}$  \\
3       &27309             &56\,Tau               &A0 SiCr     &$~~90(2)$               &$12.0 (0.5)$   &${2.3 (0.3)^{(10)}}$  &1.56882(5)$^{(18)}$          & $3.6^{+1.98}_{-0.58}$$^{(10)}$     &$0.41 (0.33 )^{(10)}$        & $3.6(0.6)\times 10^{15}$    \\
4       &34452             &IQ\,Aur               &B9 Si       &$~~145(4)~~$            &$14.1 (0.5)$   &$3.1 (0.3)^{(5)}$  &2.466264(5)$^{(19)}$         & $\approx4^{(34)}$       &$0.40^{+0.15}_{-0.21}$$^{(37)}$      & $1.0(0.3)\times 10^{16}$    \\
5       &35298             &V1156\,Ori            &B6 He-wk    &$~~371(9)~~ $           &$15.8 (0.8)$   &$2.42(0.09)^{(6)}$  &1.85458(3)$^{(20)}$          &$11.2\pm1^{(6)}$       &$0.06 (0.02 )^{(6)}$        & $~~~~~~4(2)\times 10^{16}$  \\
6       &35502             &BD-02\,1241           &B5V He-wk   &$383(9)$                &$18.4 (0.6)$   &$3.0 (0.1)^{(6)}$  &0.853807(3)$^{(7)}$         & $7.3\pm0.5^{(6)}$        &$0.11 (0.03 )^{(6)}$      & $~~~~~~4(1)\times 10^{17}$   \\
7       &36485$^{\ast}$    &$\delta$\,Ori\,C      &B3Vp He-s   &$~~390(10)$             &$20.0 (2.0)$   &$2.97(0.08)^{(6)}$  &1.47775(4)$^{(21)}$          & $8.9\pm0.2^{(6)}$    &$0.05 (0.02 )^{(6)}$          & $1.7(0.6)\times 10^{17}$    \\
8       &37017$^{\ast}$    &V1046\,Ori            &B2 He-s     &~~~~$420(35)^{b}$       &$21.0 (2.0)$   &$3.6 (0.2 )^{(6)}$  &0.901186(2)$^{(20)}$         & $6.2\pm0.9^{(6)}$      &$0.13 (0.03 )^{(6)}$        & $~~~~~~4(2)\times 10^{17}$  \\
9       &37479$^{\ast}$    &$\sigma$\,Ori\,E      &B2 Vp He-s  &~~$440(20)$             &$23.0 (2.0)$   &$3.39^{+0.06}_{-0.04}$$^{(6)}$               &1.190833(3)$^{(22)}$    &$10.5\pm1.5^{(4)}$   &$0.06 ^{+0.02}_{-0.03}$$^{(6)}$   & $~~~~~~7(3)\times 10^{17}$  \\
10      &40312             &$\theta$\,Aur         &A0 Si       &$~~58(1)$               &$10.2 (0.4)$   &$4.6 (0.2 )^{(8)}$  &3.61866(1)$^{(23)}$          & $0.69\pm0.01 ^{(8)}$    &$1.00 ^{+0.01}_{-0.06}$$^{(10)}$       & $1.2(0.6)\times 10^{15}$    \\
11      &79158             &36\,Lyn               &B9 He-wk    &$175(6)$                &$13.1 (0.7)$   &$3.4 (0.7 )^{(9)}$  &3.83476(4)$^{(24)}$          & $3.57\pm0.36^{(9)}$      &$0.84^{+0.08}_{-0.1}$$^{(37)}$      & $1.6(0.8)\times 10^{16}$    \\
12      &112413            &$\alpha^2$\,CVn       &A0 EuSiCr   &$~~35(1)$               &$11.5 (0.5)$   &$2.5 (0.3 )^{(10)}$  &5.46913(8)$^{(16)}$          & $3.46^{+2.29}_{-0.69}$$^{(16)}$ &$0.53 ^{+0.28}_{-0.31}$$^{(10)}$     & $1.1(0.7)\times 10^{15}$  \\
13      &118022            &78\,Vir$^{\dag\dag}$      &A2 CrEuSr   &$~58.3(0.8)$   &$~~9.4(0.5)$   &$2.0 (0.1 )^{(10)}$  &3.722084(2)$^{(25)}$    & $3.65^{+0.61}_{-0.37}$$^{(16)}$  &$0.39 ^{+0.32}_{-0.25}$$^{(10)}$  & $~~~~~~2(1)\times 10^{15}$  \\
14      &124224$^{\ast}$   &CU\,Vir               &B9 Si       &$~~72(1)$               &$12.2 (0.5)$   &$2.1 (0.1 )^{(11)}$  &0.520714(1)$^{(26)}$         & $3.8\pm0.2^{(11)}$        &$0.13 ^{+0.42}_{-0.09}$$^{(10)}$     & $1.8(0.6)\times 10^{16}$    \\
15      &133652            &HZ\,Lup               &B9 Si       &$121(1)$                &$12.8 (0.5)$   &$2.2 (0.2 )^{(12)}$  &2.30405(2)$^{(12)}$          & $6.1\pm0.7^{(12)}$        &$0.06 (0.02 )^{(38)}$     & $~~~~~~4(2)\times 10^{15}$  \\
16      &133880$^{\ast}$   &HR\,5624              &B9 Si       &$104(3)$                &$12.0 (0.5)$   &$2.0 (0.3 )^{(13)}$  &0.877476(9)$^{(13)}$         & $9.6\pm1^{(13)}$     &$0.05 (0.02 )^{(38)}$        & $3.5(1.5)\times 10^{16}$   \\
17      &142184$^{\ast}$   &HR\,5907              &B2V He-s    &$141(3)$                &$18.5 (0.5)$   &$2.8 (0.1 )^{(6)}$  &0.508276(15)$^{(27)}$        & $9\pm2^{(6)}$        &$0.05_{-0.05}^{+0.1}$$^{(6)}$     & $~~~~~~2(1)\times 10^{18}$  \\
18      &142301$^{\ast}$   &3\,Sco                &B8 He-wk Si &$148(3)$                &$16.2 (0.7)$   &$2.42(0.02)^{(4)}$  &1.45957(5)$^{(28)}$          &$12.5^{+9}_{-0.3}$$^{(4)}$       &$0.04 (0.01 )^{(38)}$       & $~~~~~~9(2)\times 10^{16}$  \\
19      &142990            &V913\,Sco             &B7 He-wk    &$146(4)$                &$17.0 (0.7)$   &$2.79(0.06)^{(6)}$  &0.978891793(6)$^{(29)}$      & $4.7\pm0.4 ^{(6)}$     &$0.06 (0.01)^{(6)}$       & $4.7(0.8)\times 10^{16}$    \\
20      &144334            &HR\,5988              &B8 He-wk    &$133(2)$                &$15.2 (0.7)$   &$2.27(0.04)^{(4)}$  &1.49499(4)$^{(4)}$          & $3.6\pm0.3^{(4)}$      &$0.04 (0.01 )^{(38)}$       & $~~~~~~7(3)\times 10^{15}$  \\
21      &145501            &$\nu$\,Sco\,C         &B9 Si       &$141(1)$                &$14.0 (0.5)$   &$2.26(0.06)^{(4)}$  &1.02648(1)$^{(4)}$          & $5.8\pm0.3^{(4)}$         &$0.04 (0.02 )^{(38)}$    & $3.8(0.7)\times 10^{15}$    \\
22      &147932$^{\ast}$   &$\rho$\,Oph\,C        &B5V$^a$     &$134(1)$                &$17.0 (1.0)^{(1)}$   &$3.3 (0.2 )^{(14)}$  &0.8639(1)$^{(30)}$       &$13.2_{-1.2}^{+2.6}$$^{(14)}$   &$0.12_{-0.03}^{+0.04}$$^{(39)}$   & $~~~~~~4(1)\times 10^{17}$  \\
23      &147933$^{\ast}$   &$\rho$\,Oph\,A        &B2V$^a$     &$140(4)$                &$20.8 (0.5)^{(2)}$   &$4.5 (0.6 )^{(15)}$  &0.747326(2)$^{(2)}$     & $2.7_{-0.7}^{+0.9}$$^{(2)}$   &$0.26 (0.06) ^{(39)}$    & $1.7(0.5)\times 10^{17}$    \\
24      &170000            &$\phi$\,Dra           &A0 Si       &$~~~~93(3)^{c}$         &$11.4 (0.4)$   &$3.7 (0.1 )^{(10)}$  &1.71665(9)$^{(18)}$       & $1.75^{+0.14}_{-0.16}$$^{(16)}$    &$0.88 ^{+0.09}_{-0.03}$$^{(10)}$       & $~~~~~~6(2)\times 10^{15}$    \\
25      &175362            &HR\,7129             &B6 He-wk    &$153(4)$                &$16.9 (0.7)$   &$2.7(0.2)^{(6)}$  &3.67381(1)$^{(20)}$          & $17.0^{+0.6}_{-0.4}$$^{(6)}$        &$0.06_{-0.06}^{+0.11}$$^{(6)}$      & $~~~~~~9(3)\times 10^{15}$  \\
26      &176582            &V545\,Lyr             &B5 He-wk    &$301(4)$                &$17.2 (0.7)$   &$3.21(0.06)^{(6)}$  &1.58221(5)$^{(18)}$          & $5.4\pm0.2^{(6)}$                    &$0.32 (0.06)^{(6)}$       & $~~~~~~4(1)\times 10^{16}$  \\
27      &182180$^{\ast}$   &HR\,7355              &B2Vn        &$233(6)$                &$19.8 (1.4)$   &$3.2(0.1)^{(6)}$  &0.5214404(6)$^{(31)}$        & $9.5\pm0.6^{(6)}$                    &$0.09 (0.09)^{(6)}$       & $~~~~~~9(3)\times 10^{17}$  \\
28      &215441            &GL\,Lac$^{\dag\dag\dag}$  &B9 Si       &$~~490(10)$             &$14.7 (0.5)$   &${3 (1)}^{(40)}$  &9.487574(3)$^{(32)}$         &$\approx62.4^{(35)}$      &$0.2 (0.2)^{(40)}$        & $~~~~~~3(1)\times 10^{17}$  \\
\hline
\end{tabular}
\begin{list}{}{}
\item[Notes:] 
$^{\ast}$ Radio spectrum available.
$^{\dag}$ The uncertainties in the least significant digit of the rotation are given in parentheses.
$^{\dag\dag}$ First magnetic star \citep{babcock47}, except the Sun.
$^{\dag\dag\dag}$ Babcock's star (mean field $\approx 34$ kG; \citealp{babcock60}).
$^a$ Spectral type from Simbad. 
$^b$ HD\,37017 assumed at the average distance of the stars belonging to the Ori\,OB1c cluster \citep{landstreet_etal07},
see discussion reported in \citet{shultz_etal19_485}.
$^c$ Distance from Hipparcos \citep{hipparcos}. 
\item[References:]
$^{(1)}$\citealp{alecian_etal14}; 
$^{(2)}$\citealp{leto_etal20}; 
$^{(3)}$\citealp{allende_prieto99}; 
$^{(4)}$\citealp{shultz_etal20};
$^{(5)}$\citealp{pasinetti_etal01};
$^{(6)}$\citealp{shultz_etal19_490}; 
$^{(7)}$\citealp{sikora_etal16}; 
$^{(8)}$\citealp{kochukhov_etal19}; 
$^{(9)}$\citealp{wade_etal06}; 
$^{(10)}$\citealp{sikora_etal19a}; 
$^{(11)}$\citealp{kochukhov_etal14}; 
$^{(12)}$\citealp{bailey_etal15}; 
$^{(13)}$\citealp{bailey_etal12}; 
$^{(14)}$\citealp{leto_etal20b}; 
$^{(15)}$\citealp{pillitteri_etal18}; 
$^{(16)}$\citealp{sikora_etal19b};
$^{(17)}$\citealp{borra_landstreet80}; 
$^{(18)}$\citealp{bernhard_etal20}; 
$^{(19)}$\citealp{bohlender_etal93}; 
$^{(20)}$\citealp{shultz_etal18};
$^{(21)}$\citealp{leone_etal10};
$^{(22)}$\citealp{townsend_etal10};
$^{(23)}$\citealp{krticka_etal15};
$^{(24)}$\citealp{oksala_etal18};
$^{(25)}$\citealp{catalano_leone94};
$^{(26)}$\citealp{pyper_etal13};
$^{(27)}$\citealp{grunhut_etal12};
$^{(28)}$\citealp{shore_etal04};
$^{(29)}$\citealp{shultz_etal19_486}; 
$^{(30)}$\citealp{rebull_etal18}; 
$^{(31)}$\citealp{rivinius_etal13}; 
$^{(32)}$\citealp{north_adelman95};
$^{(33)}$\citealp{glagolevskij_gerth10};
$^{(34)}$\citealp{babel_montmerle97};
$^{(35)}$\citealp{landstreet_mathys00};
$^{(36)}$\citealp{wraight_etal12};
$^{(37)}$\citealp{kochukhov_bagnulo06};
$^{(38)}$\citealp{landstreet_etal07};
$^{(39)}$ \citealp{schneider_etal14};
$^{(40)}$ \citealp{wade97}. 
\end{list}

\end{table*}

{Significant non-thermal radio emission is dependent on the
presence of a strong magnetic field in the plasma environment of
early-type magnetic stars.}
Indeed, about 25\% of the magnetic B/A
type stars are non-thermal radio sources
\citep{drake_etal87,linsky_etal92,leone_etal94}. The prevailing
scenario to explain their radio emission is related to the interaction
between the radiatively driven ionized stellar wind and the ambient
magnetic field \citep{andre_etal88}. The ionized stellar wind opens 
the magnetic field lines, and a continuous acceleration occurs in the 
current sheets in the equatorial plane.  The 
distance at which the radiative wind kinetic energy density equals the magnetic
energy density is the Alfv\'{e}n radius ($R_{\mathrm{A}}$).
Non-thermal plasma particles, propagating within the stellar
magnetosphere, radiate in the radio regime through the incoherent
gyro-synchrotron mechanism and produce continuum radio emission
that is partially circularly polarized. 

The main observational features of non-thermal radio emission from 
early-type magnetic stars are nearly flat spectra, and 
rotational modulation, both for the total intensity (Stokes\,$I$) 
and circular polarization (Stokes\,$V$). The polarization level 
for incoherent non-thermal radio emission increases with
frequency, reaching levels of $\approx 10$--20\% 
\citep{leto_etal12,leto_etal18,leto_etal19,leto_etal20}. 
Multifrequency radio light curves for Stokes\,$I$ and $V$ have been 
successfully reproduced using a synthetic 3D model of 
gyro-synchrotron emission from an oblique dipole field and co-rotating 
magnetosphere \citep{trigilio_etal04,leto_etal06}.

Superimposed on the incoherent radio emission, highly polarized 
(up to $100$\%) strong pulses of coherent emission have also been detected at 
low frequencies in seven early-type magnetic stars
\citep{trigilio_etal00,das_etal18,das_etal19a,das_etal19, 
leto_etal19,leto_etal20,leto_etal20b}. This amplified pulsed 
emission is stimulated via the coherent Electron Cyclotron Maser 
(ECM) emission mechanism \citep{wu_lee79,melrose_dulk82}. ECM 
is also the process explaining highly beamed auroral radio emissions 
(AREs) from the magnetized planets of the solar system \citep{zarka98}.
In Jupiter, 
auroral X-rays \citep{branduardi-raymon_etal07,branduardi-raymon_etal08} 
arise from the footprints of the
magnetic field lines where the coherent auroral radio emission
originates, at a few stellar radii above the magnetic poles.
ARE from some early-type magnetic stars also has 
corresponding X-ray emission
\citep{pillitteri_etal16,pillitteri_etal17,robrade_etal18}.

Interestingly, in the lower main sequence, signatures of coherent 
AREs have also been identified in ultra-cool dwarf stars and 
brown dwarfs (UCDs) \citep{berger02,kao_etal18}. 
These cold 
stars are fully convective, yet a well-ordered dipole-dominated 
magnetic field topology has been recognized in many of them 
\citep{donati06,morin_etal10}. The magnetic fields of these stars 
are generated by a 
rotational-convective dynamo analogous to the Earth's geodynamo
\citep{christensen_etal09}. 
It is worth 
noting that in some cases rotationally modulated incoherent 
non-thermal emission has also been detected 
\citep*{mclean_etal11,williams_etal15}.

The aim of this paper is to test the scenario capable of explaining 
the radio emission of hot magnetic stars, that works for 
early B-type magnetic stars, through to magnetic stars at the 
low-temperature boundary of this group (late B/early A). For this
purpose, we have identified a sample of B/A type magnetic stars 
that have been detected at radio wavelengths, and have collected
reliable information about their stellar parameters. The stellar sample is
described in Section~\ref{sec_sample}. Details regarding the radio
measurements of the individual stars are reported in
Appendix~\ref{radio_appendix}. For some stars we collected enough
multifrequency radio measurements to produce reliable radio spectra.
Model {calculations} 
of the radio spectra of this representative 
sub-sample are discussed in Section~\ref{sec_subsample}. The results
of the modeling raised a critical issue regarding the assumed wind 
paradigm that, up to now, has been assumed as a primary engine to support
and explain all the observed radio phenomena from early-type 
magnetic stars (we discuss this in Section~\ref{sec_wind}). 
We have searched for a 
possible dependence of the radio luminosity with stellar age in 
Section~\ref{sec_age}. By correlating the radio luminosities of 
the whole sample of stars with the parameters of their stellar 
magnetospheres, we obtained an empirical relation between the 
incoherent non-thermal radio luminosity of the B/A magnetic stars 
and the combined effect of rotation and magnetic properties of their 
large-scale dipole-dominated magnetospheres, which we describe in 
Section~\ref{sec_rotation}. In Section~\ref{other_cases}, this new 
result is applied in a broader context involving other classes 
of stars (and planets) that radiate incoherent non-thermal radio 
emission and have large scale dipole dominated magnetospheres, such 
as Jupiter and the UCDs.
{In Sec.~\ref{hot_jupiters}, the scaling relationship
for the incoherent non-thermal radio emission has been applied to a small sample of magnetic hot
Jupiters to estimate the expected flux levels and the possible detectability.}
In Section~\ref{discussion} we discuss the possible 
physical consequences of our results, and in Section~\ref{sec_conclusion} 
we present our conclusions.

\begin{figure*}
\resizebox{\hsize}{!}{\includegraphics{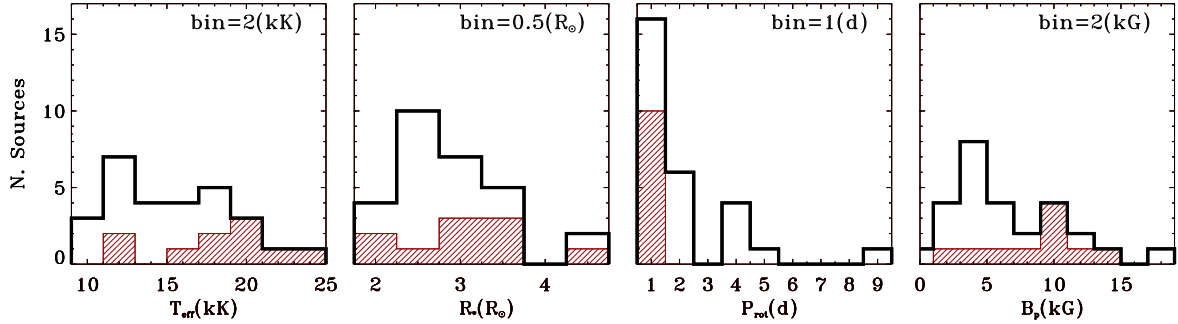}}
\caption{Parameter distributions of the analyzed sample. 
The sub-sample analyzed in Sec.~\ref{sec_subsample}
is shown by the colored areas.
Note that the Babcock's star lies outside the range of the $B_{\mathrm p}$ plot at 62.4 kG.}
\label{isto4}
\end{figure*}

\section{The sample}
\label{sec_sample}

Our sample consists of stars 
that satisfy two conditions: each star has at least one
observation in which radio emission was clearly detected, and 
{stellar and magnetospheric parameters are available for each.}
The selected sources and stellar parameters 
with the corresponding references, are listed in Table~\ref{tab_stars}.  
Each source is identified by the number listed in column 1. Common 
and alternative names are given in columns 2 and 3. Column 4 lists 
the spectral types and chemical peculiarities of the selected stars, 
which together with the effective temperature (column 6), are mainly 
taken from \citet{netopil_etal17} and \citet{shultz_etal19_485}.  
Column 5 reports the distances of the sources from \citet{gaia_dr2};
in cases when Gaia measurements are not available, references
for adopted distances are identified. Radii, rotation periods, and
polar magnetic field strengths are given in columns 7, 8 and 9.
Column 10 lists the fractional MS ages retrieved from the literature.  
In two cases the stellar evolutionary state had not previously 
analyzed, so for these 2 stars (HD\,147932 and HD\,147933)
we used BONNSAI\footnote{The BONNSAI web-service is available at 
\url{www.astro.uni-bonn.de/stars/bonnsai}} \citep{schneider_etal14} 
to estimate ages using the evolutionary models given by \citet{brott_etal11}.  
Column 11 lists the average radio luminosities 
of each star obtained from averaging all available radio measurements.

Radio measurements of individual stars have been collected from
both published measurements and unpublished observations, acquired 
by us or retrieved from the VLA data archive. The archival 
data have been analyzed using the standard reduction steps enabled 
within the software package {\sc casa}. We also included data from 
the New VLA Sky Survey (VLASS) \citep{lacy_etal20}. The VLASS 
observed the sky at $\delta>-40^{\circ}$ and at $\nu=3$ GHz (2 GHz bandwidth). The fluxes have been corrected 
for systematic errors as reported by \citet{lacy_etal19}. 
For the newly obtained radio measurements,
we determined fluxes using a gaussian fit to the radio sources (using 
the standard procedure in {\sc casa}) located at the sky position of the stars in our sample.
Details regarding the radio measurements for individual stars are provided 
in Appendix~\ref{radio_appendix}, where the mean (or unique) radio 
frequencies of all the available flux measurements are also given.
Representative frequencies where the stellar radio luminosities 
have been estimated lie in the range $\approx 3$--13 GHz.

The stellar parameters of our sample ($T_{\mathrm{eff}}$,
$R_{\ast}$, $P_{\mathrm{rot}}$, $B_{\mathrm p}$) and the corresponding
distributions are shown in
Fig.~\ref{isto4}, where the distributions of the sub-sample 
analyzed in detail in Sec.~\ref{sec_subsample} are also superimposed.  
The effective temperatures of the selected stars ranges from $\approx 10$ 
kK up to near $\approx 25$ kK, with a broad distribution roughly
peaked close to $T_{\mathrm{eff}} \approx 15$ kK.  The radii of the
selected stars cover the range 2--5 R$_{\odot}$.  The stellar 
rotation periods of the sample are mainly between 
$\approx 0.5$ and $\approx 2.5$ days, with a secondary group close 
to $P_{\mathrm{rot}} \approx 3.5$--4 d.  Only two targets have 
longer rotation periods (third panel of Fig.~\ref{isto4}). Finally, 
the last panel of Fig.~\ref{isto4} shows the polar magnetic field 
strength distribution. The stars analyzed here have polar magnetic 
field strengths peaked close to $B_{\mathrm p} \approx 3$--4 kG,
with a clear secondary peak centered at $B_{\mathrm p} \approx
9$--10 kG.  Note that Babcock's star ($B_{\mathrm p}=62.4$ kG) is
not shown in the fourth panel of Fig.~\ref{isto4} to avoid
enlarging the x-axis scale for just this one source.

\begin{figure*}
\resizebox{\hsize}{!}{\includegraphics{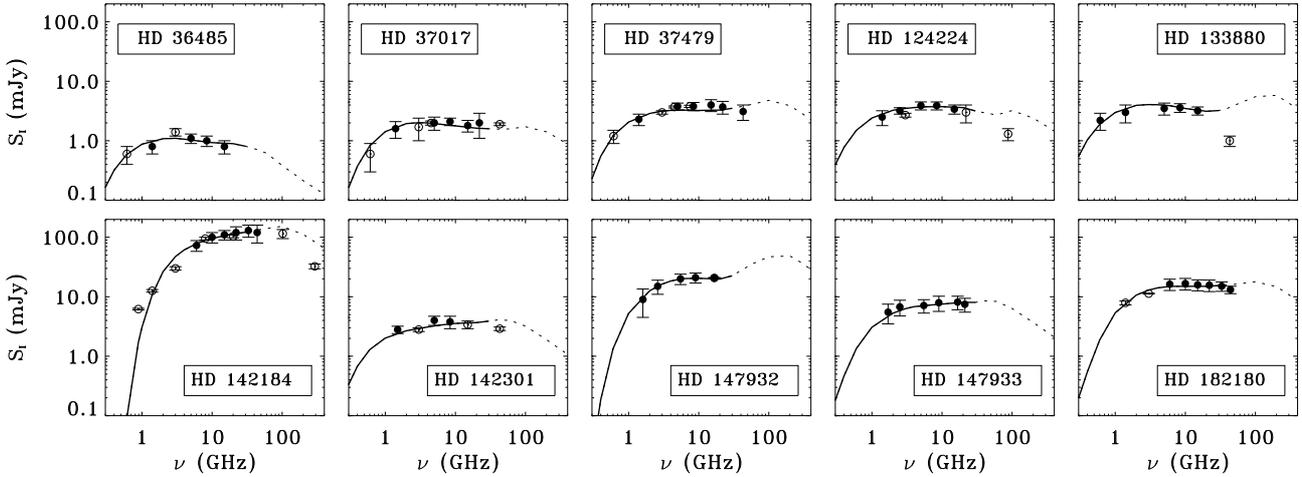}}
\caption{Radio spectra of the ten stars having enough radio measurements
to produce reliable average radio spectra. 
The data are provided in
the tables reported in Appendix~\ref{radio_appendix}, where details
on the individual stars are also described.
Open symbols refer to observing frequencies having few measurements (less than three);
in these cases the sampling of the stellar rotation period is poor. 
The modeled radio spectra are superimposed on the observed data.
The dotted line show the model extrapolation to
higher frequencies.
}
\label{spe10sou}
\end{figure*}

\section{Modeling of a representative sub-sample}
\label{sec_subsample}

Incoherent gyro-synchrotron emission from the magnetospheres
of magnetic B/A stars is rotationally modulated
as a consequence of the variable projected source area.
To conduct a comprehensive study of radio variability ideally 
multi-frequency radio observations that
sample the entire rotation period are needed. A detailed study of the temporal 
evolution of the broadband radio spectra has been 
performed only for a few stars, and limited only to narrow ranges 
of rotational phases \citep{leto_etal12,leto_etal17,leto_etal18,leto_etal20}. 

We collected all available radio measurements for the stars in our sample
(Sec.~\ref{sec_sample}). Unfortunately
there are many cases in which the data have inadequate coverage of 
stellar rotational phase, or only a few radio frequencies available.  
Even so, a representative number of stars ($\approx 35\%$ of the 
total) do have a large number of multi-frequency radio {measurements;
their parameter distributions are also highlighted by the colored areas in Fig.~\ref{isto4}.
Such sub-sample of}
stars with superior frequency coverage
(typically, repeated measurements of at least 5 radio frequencies)
{are listed in Table~\ref{tab_stars_spectra}.}
Ten stars span nearly the whole range of considered stellar parameters, 
except for the rotation period. All
the stars of this sub-sample have short periods (less than 1.5~d).
This is an expected bias because only a few observing facilities are
able to detect stellar radio emission, and the ability to perform
radio measurements covering large portions of the stellar rotation
is naturally biased toward short-period cases.
For this sub-sample of ten stars, we obtained reliable radio spectra
from averaging observations repeated over more than 3 epochs acquired
at the same frequency. 
The average spectra of these stars are shown
in Fig.~\ref{spe10sou}.

\subsection{Model description}
\label{sec_model}

\begin{figure}
\resizebox{\hsize}{!}{\includegraphics{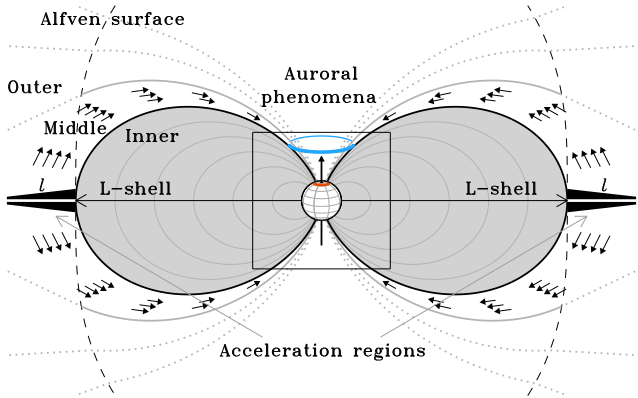}}
\caption{Meridional cross section of the dipolar dominated magnetosphere
of a typical hot magnetic star.  Beyond the
Alfv\'{e}n surface the magnetic field strength is not able to
confine the ionized plasma pushed by the radiatively driven stellar
wind (outer magnetosphere), with consequent formation of an
equatorial magneto-disk surrounding the stellar magnetosphere,
shaped like a mashed donut with a longitudinal thickness $l$.
Closer to the star, the magnetic field is strong enough to
confine the ionized gas and force it to co-rotate with the
star (inner magnetosphere). The two magnetic field lines, pictured
using solid lines (black: inner boundary; grey: outer
boundary), locate the magnetospheric cavity (middle magnetosphere)
where non-thermal electrons {(evidenced by the small arrows)}, 
accelerated up to relativistic
energies in the equatorial magneto-disk (black regions), produce
radio emission by the incoherent gyro-synchrotron mechanism.
Outside the Alfv\'{e}n surface, the magnetic field topology (dotted
lines) cannot easily be described by simple dipolar field lines.
The box centered close to the star locates the deep magnetospheric
regions where auroral phenomena can occur, including the coherent
cyclotron maser (blue ring a few stellar radii above the poles) and
the non-thermal X-rays from the auroral polar caps on the stellar
surface (orange ring).}
\label{scenario}
\end{figure}

The 3D model of stellar radio emission from a dipole dominated 
magnetosphere \citep{trigilio_etal04,leto_etal06} has been previously used to 
successfully reproduce the multi-wavelength radio light curves of 
early-type magnetic stars, both for the total intensity 
(Stokes\,$I$) and for circularly polarized emission (Stokes\,$V$).
The stellar geometry is described by the two characteristic angles
of the ORM: the rotation axis inclination ($i$) from the line of sight and the tilt angle
of the magnetic dipole axis ($\beta$) from rotation axis.
The scenario that outlines this model is visualized in Fig.~\ref{scenario}.

The calculation of observable properties involves a number of
ingredients as described below. We use a 3D cartesian grid to sample the
space around the star. Physical parameters that are functions of the
radial distance from the star are discretely sampled with the cubic
volume elements of the grid, and in each small volume element, the
physical parameters are considered constant.

In the case of a simple dipolar magnetic topology, the magnetic field 
vector components of each grid point have been calculated in the 
reference frame fixed to the magnetic dipole and then rotated to 
the observer reference frame. Consequently a distant observer will
perceive the vector magnetic field to change as the star rotates
(see Appendix of \citealp{trigilio_etal04}).

The acceleration of relativistic electrons is assumed to coincide
with the magneto-disk, where magnetic reconnection events are 
expected to occur. This region is described as a thin equatorial 
disk of length $l$ and located beyond the last closed magnetic field 
line, corresponding to the distance $L$ where the last closed
magnetic field line crosses the magnetic equatorial plane. 
It is expected that the magnetic field topology of this {``unstable''} magnetospheric region 
cannot be easily described with simple dipole magnetic field lines.
However, {we empirically observed that the contribution to the radio emission of 
the magnetospheric regions located far from the star and close to the magnetic equator
is negligible, particularly at the intermediate and high frequency ranges (Sec.~\ref{sec_spectra_behavior}).
The non-dipolar topology of the equatorial magneto-disk has then
negligible effects on the calculation 
for non-thermal radio emission.}
In fact, the bulk of the non-thermal radio emission originates from 
magnetospheric regions at higher magnetic latitudes, regions that 
are likely well-described by the simple dipole topology.
Hence, the dipole like magnetic shell where the relativistic electrons 
freely propagate is delimited by the two magnetic field lines of $L$ and $L+l$.

In modeling the non-thermal emission, we take the relativistic
electron population to conform with an isotropic pitch-angle
distribution (angle $\phi$ formed by the velocity direction with
respect to the local magnetic field vector) and having a power-law
energy distribution ($\propto E ^ {-\delta}$) with spectral index
$\delta$. Up to the locations where the non-thermal emission is
produced, the non-thermal electrons 
are assumed to freely propagate in vacuum, 
in accordance with the low density of the wind electrons
(order of magnitude $10^6$ cm$^{-3}$ at the Alfv\'{e}n radius), 
so the collisional energy loss mechanism is then negligible \citep{petrosian85}.

In the middle magnetosphere, the number density of
non-thermal electrons is approximately constant. This arises because
of two effects. First, the invariance of the magnetic moment of
the gyrating electron ($\propto \sin^2 \phi /B$) causes the magnetic
mirroring effect on the electrons moving towards regions of increasing
magnetic field strength, which results in a decreasing number of
non-thermal electrons able to reach the deep magnetospheric regions.
On the other hand, the solenoidal condition of the magnetic field
causes a decreasing volume of the dipole-like flux tube as the
magnetic field strength increases, balancing the decreasing number of the non-thermal electron
when approaching the stellar surface. The result is a
nearly constant number density ($n_{\mathrm r}$) of the non-thermal
electrons responsible for the radio emission.

All the physical parameters needed for the calculation of the
emission and absorptions coefficients for the gyro-synchrotron
emission mechanism \citep{ramaty69,klein87} are calculated within
each grid point (i.e. the local magnetic field strength and its orientation).
As the final step, before the integration of the
radiative transfer equation, the spatial distribution of the thermal
electron density trapped within the inner magnetosphere is estimated
using the MCWS model, which assumes the density is linearly decreasing
outward while the temperature is linearly increasing. 
{The simple relationship: $T \propto r$ ($r$ radial distance),
was deduced by \citet{babel_montmerle97} via 
modeling of the post-shock region 
in the case of fast rotating stars.
Further, following the MCWS model,
the thermal plasma pressure plus the wind ram pressure is constant.
Far from the boundary shock the wind pressure becomes negligible, then 
in the post-shock region $n(r) T(r) = n_0 T_0$, 
following the isobaric approximation.} 
The thermal
electron density at the stellar surface ($n_0$) is a free parameter,
whereas the temperature there {($T_0$)} is set to $T_{\mathrm{eff}}$. 
Our model accounts for free-free absorption from thermal electrons
\citep{dulk85}. 
Then at a given radio frequency ($\nu$), we perform
a numerical integration of the radiative transfer equation along the
line of sight. The solution for a grid of such rays allows us to
calculate the spatial distribution of the brightness and for a given
stellar distance, the predicted flux from the unresolved source.

A model realization requires the following parameters: the equatorial
radius of the magnetic shell where the relativistic electrons freely
propagate ($L$); the thickness of the magnetic shell ($l$); the
spectral index for the energy distribution of relativistic electrons
($\delta$); the number density of the relativistic electrons
($n_\mathrm{r}$); and for the thermal electrons trapped within the inner
magnetosphere, the number density at the stellar surface ($n_0$).
Note that \citet{leto_etal17} showed that $n_\mathrm{r}$ and $l$
are degenerate parameters, so that only the
column density of the relativistic electrons ($n_\mathrm{r} \times l$)
can be constrained.

\subsection{Radio spectra calculation over broad spectral range}
\label{sec_spectra_sim}

For each of the ten stars in the sub-sample,
we {calculated} the non-thermal radio emission
within the frequency range of 0.1--400 GHz.  We adopted a spacing
of $\Delta \log \nu= 0.2$, corresponding to 19 different frequencies
in the {calculations},  
to evaluate time-averaged synthetic spectra.  In
particular, we analyzed the orientations corresponding 
to the two extrema of the effective magnetic field curves and to one null
(if predicted by the corresponding ORM geometry, otherwise the intermediate phase 
between the extrema was {calculated}), 
for a total of 3 different rotational phases.  These particular phases
were appropriately identified from each individual star using the
ORM parameters as listed in Table~\ref{tab_stars_spectra} (thus
each star has 3 tailored {calculations} appropriate for the 3 phases identified).

To make this large computational effort possible, we restricted the
exploration of parameter space to $L$ and $n_\mathrm{r} \times
l$.  The other parameters of the models were assumed close to the values
retrieved by analyzing the individual stars already published 
\citep{trigilio_etal04,leto_etal06,leto_etal17,leto_etal18,leto_etal20,leto_etal20b},
having well-sampled single-frequency radio light curves, both for
Stokes\,$I$ and $V$.  In those cases, a fine-tuning of all the model
parameters was performed to reproduce the observed light curve
shapes.  The previously published models have thermal electron density
numbers determined at the stellar surface in the range 1--$5\times
10^9$ cm$^{-3}$ and spectral indices of the non-thermal electron
energy distribution in the range 2--4.  These values are fairly
similar for all the previously analyzed stars.  Consequently,
for this paper we adopt: 
$n_0=3 \times
10^9$ cm$^{-3}$ and $\delta=2.6$.  For $n_0$ we used the average of
the existing ranges of values.  The adopted $\delta$ is the average
of the values indirectly estimated for the few stars for which their
X-ray spectra displayed a non-thermal photon component
\citep{leto_etal17,leto_etal18,pillitteri_etal18,robrade_etal18}.

To perform the broadband spectral analysis, the radio spectral
{calculations}
were obtained by varying the parameter $L$ in the range 
8--20 R$_{\ast}$ (with steps of $\Delta L= 0.5$ R$_{\ast}$,
corresponding to 24 different magnetospheric sizes) and $n_\mathrm{r} 
\times l$ in the range $10^{15}$--$10^{17}$ cm$^{-2}$ (with steps 
of $\Delta \log (n_\mathrm{r} \times l)=0.05$, corresponding to 40 
different cases). The total number of {calculations} 
is $\approx 5.5 
\times 10^5$. The spatial grid has different levels of resolution 
depending on its proximity to the star. We used narrow steps (0.1 
R$_{\ast}$) for distances below 8 R$_{\ast}$; mid-sized steps (0.3 
R$_{\ast}$) for intermediate distances in the range 8--12 R$_{\ast}$; 
and finally coarse steps (0.5 R$_{\ast}$) at distances beyond 12 R$_{\ast}$.

\begin{table}
\begin{center}
\caption[ ]{Magnetosphere parameters used to {calculate} 
the stellar radio spectra shown in Fig.~\ref{spe10sou}.
Fixed parameters: $n_0 = 3 \times 10^{9}$ cm$^{-3}$; $\delta=2.5$.
}
\label{tab_stars_spectra}
\footnotesize
\begin{tabular}{l @{~~~}c @{~~~}c @{~~~}c @{~~~}c c @{~~~}c @{~~~}c}

\hline
Star                 &$i$               &$\beta$      &$L$                   &$B_{L}$      &{$\dot{M}_L$}                 &$n_{\mathrm r} \times l$       &$\tilde{\chi}^2$   \\
 HD                  &{(deg)}            &{(deg)}           &{(R$_{\ast}$)}      &{(G)}        &{(M$_\odot$\,yr$^{-1}$)}   & {($10^{15}$\,cm$^{-2}$)}      &    \\
\hline       
36485$^{a}$  &$19$             &$~~4$         &$16$                    &$1.1$   &{$~~5.1 \cdot 10^{-10}$}               &$3.5$                        &0.6      \\  
37017$^{b}$ &$39$            &$57$            &$13$                     &$1.4$   &{$~~5.9 \cdot 10^{-10}$}                &$10$                  &{0.7}       \\
37479$^{b}$ &$77$             &$38$            &$15$                   &$1.6$    &{$1.4 \cdot10^{-9}$}              &{11.2}     &0.6      \\ 
124224$^{c}$  &$46$         &$76$             &$12$                     &$1.1$   &{$~~1.5\cdot 10^{-10}$}               &{1.6}       &0.8     \\ 
133880$^{d}$ &$55$             &$78$          &$18$                     &$0.8$   &{$~~2.4\cdot 10^{-10}$}               &$3.5$                            &0.3    \\ 
142184$^{b}$ &$64$            &$~~9$         &$10$                     &$4.5$   &{$1.2\cdot 10^{-9}$}                &{56}            &0.5         \\  
142301$^{a}$  &$68$              &$58$          &$19$                   &$0.9$    &{$~~4.9\cdot 10^{-10}$}             &$2.8$                         &2   \\ 
147932$^{e}$  &$74$              &$~~5$      &$13$                    &$3.0$    &{$1.3 \cdot10^{-9}$}              &{6.3}          &0.04               \\ 
147933$^{f}$   &$35$             &$78$        &$ 9$                      &$1.8$    &{$~~7.5 \cdot10^{-10}$}              &$ 2.8$          &0.1                  \\
182180$^{a}$  &$53$              &$82$        &$ 14$                   &$1.7$     &{$~~7.7\cdot 10^{-10}$}             &{11.2}               &0.2        \\  
\hline
\end{tabular}
\begin{list}{}{}
\item[]References of the stellar geometry:
$^{a}$ \citealp{shultz_etal20}; 
$^{b}$ \citealp{shultz_etal19_486}; 
$^{c}$ \citealp{kochukhov_etal14}; 
$^{d}$ \citealp{bailey_etal12}; 
$^{e}$ \citealp{leto_etal20b}; 
$^{f}$ \citealp{leto_etal20}.
\end{list}

\end{center}
\end{table}

Finally, the grid of {calculated} 
spectra was used to find the best match to observations.
For goodness of the spectral fit, 
we calculated the reduced chi-square statistical
parameter, defined as $\tilde{\chi}^2 = \chi^2/{\mathrm{d.\,o.\,f.}}$,
where the degrees of freedom (d.\,o.\,f.) is equal to the number
$n$ of the different observed frequencies minus the two free
parameters of the model. 
The formulation ${\chi}^2
= \sum_{i=0}^{n} (O_i - E_i)^2/\sigma_i^2$ has been used, where $O_i$ is
the measured flux (with the related uncertainty $\sigma_i$) and $E_i$ is the expected value from the {calculation} 
at the corresponding $i$-th frequency.  
For the evaluation of
$\tilde{\chi}^2$, the {calculated} 
spectra have been interpolated 
to match the frequencies of the observations. 
To make the $\tilde{\chi}^2$ estimation reliable,
we related to the radio measurements marked by the open symbols in Fig.~\ref{spe10sou}
(spectral data provided by unique or few measurements)
the average fractional error gained by the other measurements,
in this way the dispersion of the measurements due to 
the rotational modulation of the radio emission is taken into account. 
Further, as it will be extensively discussed later,
we suspect that the theoretical high-frequency radio emission suffers 
from critical model limitations, then, 
the high-frequency
measurements ($\nu \gtrapprox 30$ GHz), when available, have not
been taken into account when evaluating $\tilde{\chi}^2$.
The  {calculated} 
spectra superimposed with the observations
are shown in Fig.~\ref{spe10sou}. 

Table~\ref{tab_stars_spectra} lists model parameters for the synthetic
spectra that best fit the data, including the magnetic field strength
($B_L$) at $L$. 
Values of $\tilde{\chi}^2$ (Table~\ref{tab_stars_spectra})
are always lower than 1 (or close to 1 in  the case of HD\,142184).
Our new spectral  {calculations}  
recover parameters that are fairly
similar to what had been obtained for stars that had previously
been modeled in terms of rotational modulations. This is encouraging
since here we are modeling spectra whereas before the emphasis was
on light curves, and suggests that our approach produces consistent
and reliable results. Overall,  {calculations}  
for these 10 stars
indicate that on average, the relativistic electrons are injected
at a distance of $\langle L \rangle=14 \pm 3$ R$_{\ast}$, and the local magnetic
field strength is $\langle B_{L} \rangle=2 \pm 1$ G.

\subsection{The broadband radio spectra}
\label{sec_spectra_behavior}

The  {calculated}  
radio spectra at $\nu \lessapprox 30$ GHz for the ten magnetic stars in our sub-sample
are quite independent of
their stellar parameters. 
The spectral shapes 
suggest the existence of both low- and high- frequency cutoffs,
in accordance with the observations.
Early B and late A magnetic stars have almost indistinguishable radio 
spectra (Fig.~\ref{spe10sou}). The small differences 
are mainly related to the individual magnetospheric geometries. 
The nearly universal spectral behavior 
can be summarized as follows: the radio flux 
increases with increasing frequency (at low-frequency); 
then the radio spectrum becomes almost flat (at intermediate-frequency);
and finally the model spectrum has a negative slope 
(at high-frequency), although in some cases a clear 
high-frequency emission peak was predicted by our {calculations}.

To better understand the overall spectral behavior, we computed the
spatial distribution of the radio emission from a typical
dipole-dominated magnetic star. The synthetic radio maps have been
obtained at low-, intermediate-, and high-frequency ranges. The
three panels of Fig.~\ref{maps} show the {synthetic} 
radio maps for
these three radio regimes. The three
{calculated}  
frequencies are respectively: low frequency ($\nu=2$ GHz);
intermediate frequency ($\nu=6$ GHz); and high frequency ($\nu=100$ GHz).
We used HD\,142184 as a template,
with a polar field strength of $B_{\mathrm p}=9$ kG. 
The radio maps clearly highlight the spatial location within the
middle-magnetosphere of the regions that mainly contribute to the
emission at a given frequency range.

\begin{itemize}
\item[]{\bf Low frequency behavior.}   
The low-frequency side of the
radio spectrum tracks the radio emission arising from the furthest
regions of the middle-magnetosphere: see the top panel of Fig.~\ref{maps}.
As previously explained (see Sec.~\ref{sec_model}), each grid point
of the 3D model is a cube-shaped homogeneous source characterized by a
typical gyro-synchrotron radio spectrum peaked at the turnover
frequency, which is proportional to the local magnetic field strength
\citep{dulk_marsh82}.  
The positive slope of the radio spectrum at the low-frequency regime, observed and  {calculated}, 
is due to the grid points, that should have contributed mainly to the low-frequency emission,
being external to the middle-magnetosphere and are not crossed by relativistic electrons.
Hence, far from the star, the magnetospheric
regions that should mainly radiate at low-frequencies ($\nu
\lessapprox1$--2 GHz), and where the local magnetic field strength
is low (typically less than a few gauss), fall outside the
middle-magnetosphere. 
These outer regions do not contribute to the non-thermal radio emission, 
{preventing also to further investigate
for the possible effects of the non dipolar topology of the equatorial magneto-disk on the calculation of the radio emission.}

\begin{figure}
\resizebox{\hsize}{!}{\includegraphics{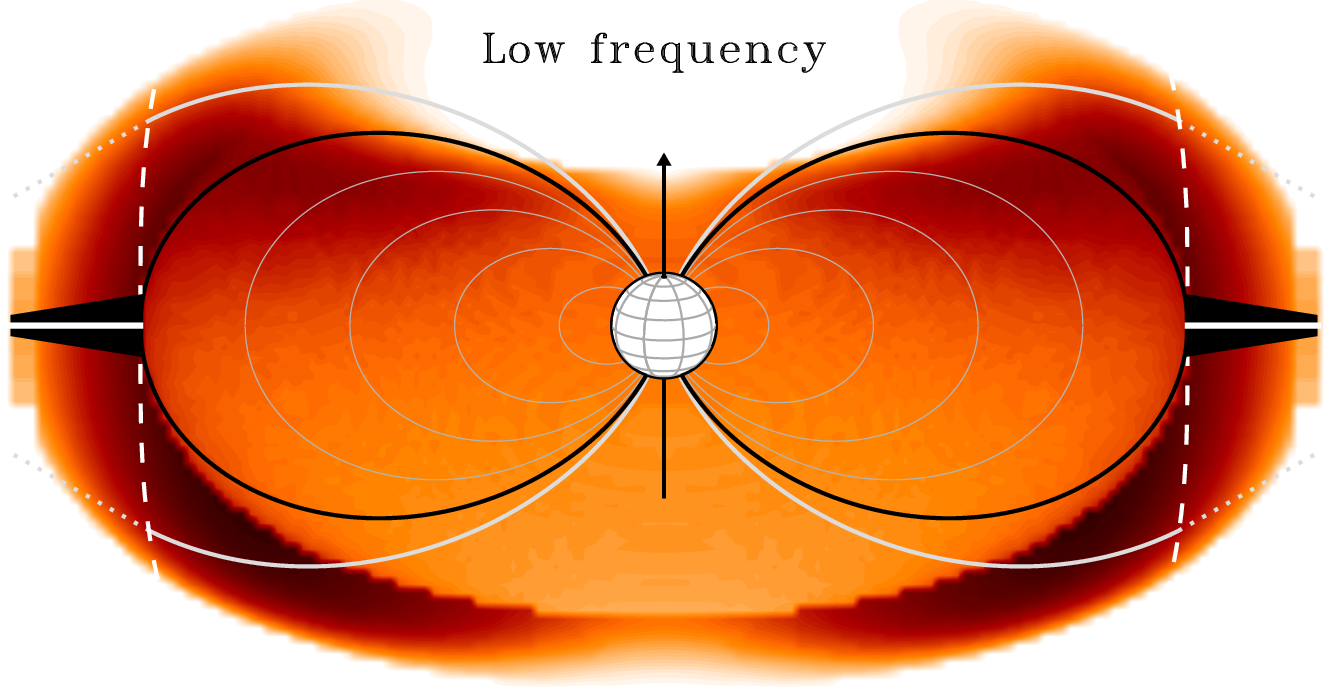}}
\resizebox{\hsize}{!}{\includegraphics{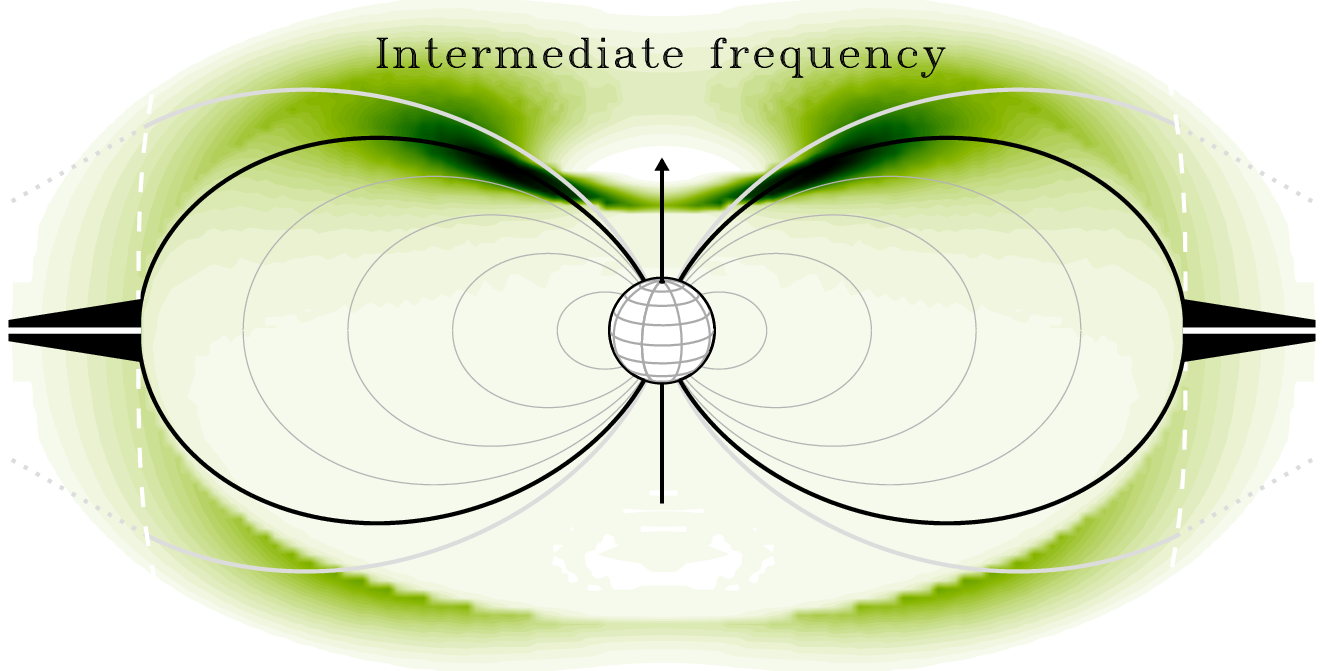}}
\resizebox{\hsize}{!}{\includegraphics{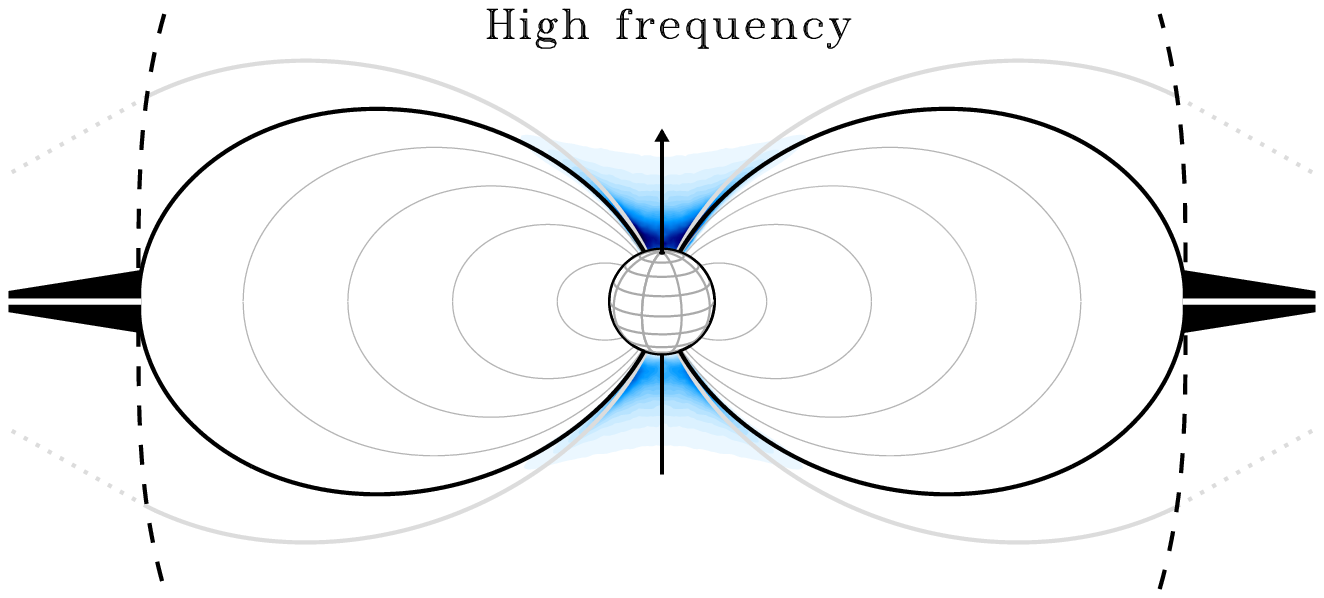}}
\caption{
Synthetic brightness spatial distributions evaluated at three representative radio frequencies
computed for the parameters of the template star HD\,142184.
Top panel: low frequency behavior (red, $\nu=2$ GHz). 
Middle panel: intermediate frequency behavior (green, $\nu=6$ GHz). 
Bottom panel: high frequency behavior (blue, $\nu=100$ GHz). 
In all panels, the dashed line demarcates the Alfv\'{e}n surface (white lines in top and middle panels, black line in the bottom panel).}
\label{maps}
\end{figure}

\item[]{\bf Intermediate frequency behavior.}
The radio emission produced at this frequency range ($ \approx 2$--30 GHz)
has the optimal tuning for non-thermal
radio emission from a dipole-dominated magnetosphere.  In fact, the
frequency peaks of the elementary spectra produced by each homogeneous
element sampling the middle magnetosphere fall within this spectral
range.  The synthetic radio map reveals that the brighter source
regions are located far from the middle-magnetosphere boundaries,
either near the stellar surface or near the Alfv\'{e}n surface (see the middle
panel of Fig.~\ref{maps}).  The nearly flat spectra, which characterize
the intermediate frequency behavior, is qualitatively understood when 
taking into account that the brighter magnetospheric regions close
to the star (where the magnetic field is stronger) have less volume than
the more distant low brightness regions, where the emission at the
lower radio frequencies mainly originates.

\item[]{\bf High frequency behavior.} 
Radio emission at higher
frequencies ($ \gtrapprox 30$ GHz) originates from the deep magnetospheric regions close
to the star (see the bottom panel of Fig.~\ref{maps}).  In this frequency
regime the negative slope of the spectrum is the obvious consequence
of the radiating regions that meet the stellar surface. 
The decaying of the
high-frequency flux has the same qualitative explanation
as the flux drop occurring when the radio frequency decreases, which
is that the magnetospheric regions that should have been mainly responsible for
radio emission at very high (and low) frequencies fall outside
the middle-magnetosphere.  Further, the high-frequency side of the
radio spectrum might also be strongly affected by the plasma processes
responsible for auroral phenomena (i.e., possible plasma
evaporation as a consequence of non-thermal auroral X-ray
emission), which is plausibly always occurring for all magnetospheres
in these kinds of stars.  This might be a critical issue for 
high frequency radio emission.  In practice, the adopted model
conditions may not be valid within deep magnetospheric regions
where high-frequency emission originates (i.e., the spatial
distribution of the non-thermal electron density might be inhomogeneous).
The topic will be a matter of future study, mainly focused on the
high-frequency side of the radio spectrum of such stars.  At present,
the limited available radio measurements prevent further investigation 
of the physical conditions occurring within the deep magnetospheric regions.  
The {calculated} 
radio light curves at the high-frequency spectral range 
are shown using only dotted lines in Fig.~\ref{spe10sou}.

\end{itemize}

The comparison between observations and  {calculations}  
shows excellent 
accord at the low and intermediate frequency ranges (Fig.~\ref{spe10sou}). 
At the higher frequency side of the radio spectra, the few available 
observations (see Appendix~\ref{radio_appendix} for details) are 
not sufficient to assert any firm conclusions regarding 
high-frequency spectral behavior, nor reliable estimates of the high 
frequency cutoff. The evident large discrepancies between  {calculations}  
and observations require further investigation. In practice, 
possible plasma effects related to auroral phenomena might 
significantly affect the emission level and the rotational modulation 
of high-frequency radio emission.  This possibility drives the
need for a much better sampling of the entire stellar rotation period
for a larger sample of stars
to determine the average high-frequency spectral behavior.

Despite some modeling limitations, particularly at high-frequency,
the capability to reproduce the spectral behavior at low and
intermediate frequencies allows us to place strong constraints on
the spatial location for the acceleration of the non-thermal
electrons.  From a qualitative point of view, once all the magnetosphere
parameters have been fixed, the radio spectrum from a large
magnetosphere (large $L$) is characterized by a turn-over, namely 
the narrow spectral region where the change from the rising to the 
flat regime of the radio spectrum occurs, situated at low frequency.  
Conversely, the radio spectrum from a smaller magnetosphere
(small $L$) has its turn-over located at a higher frequency.

\begin{figure}
\resizebox{\hsize}{!}{\includegraphics{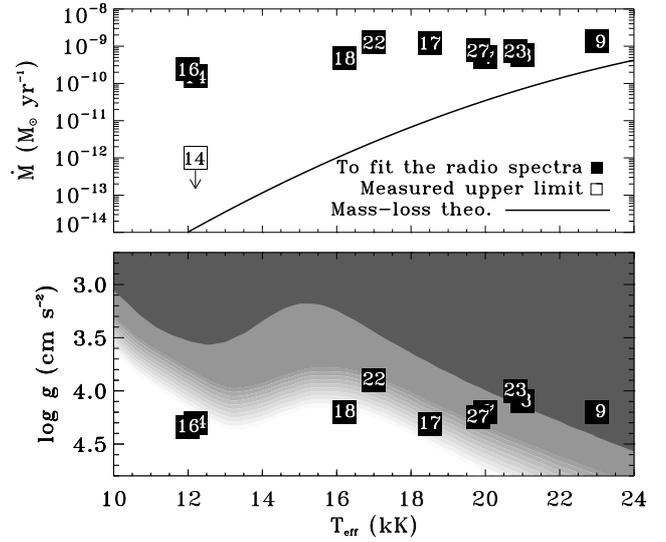}}
\caption{Top panel: theoretical wind prediction of \citet{krticka14} (solid line)
compared with 
the mass loss rate from 
modeling of the radio spectra {listed in Table~\ref{tab_stars_spectra}} (black squares). 
The numbers within squares correspond to object in Table~\ref{tab_stars}.
The open square 
represents the upper limit estimation of the wind from HD\,124224. 
Bottom panel: location of the sub-sample stars on the 
$\log g$ vs. $T_{\mathrm{eff}}$ diagram.
The {shaded} areas indicate different wind regimes, according to 
\citet{babel96} and \citet{hunger_groote99}.
The wind regions have been taken from Fig.~6 of \citet{babel96} and Fig.~1 of \citet{hunger_groote99}.
Dark gray locates the homogeneous wind region; 
light gray locates where the expected wind is inhomogeneous; 
the white area corresponds to { the static atmosphere.}
}
\label{wsim_teo}
\end{figure}

\begin{figure*}
\resizebox{\hsize}{!}{\includegraphics{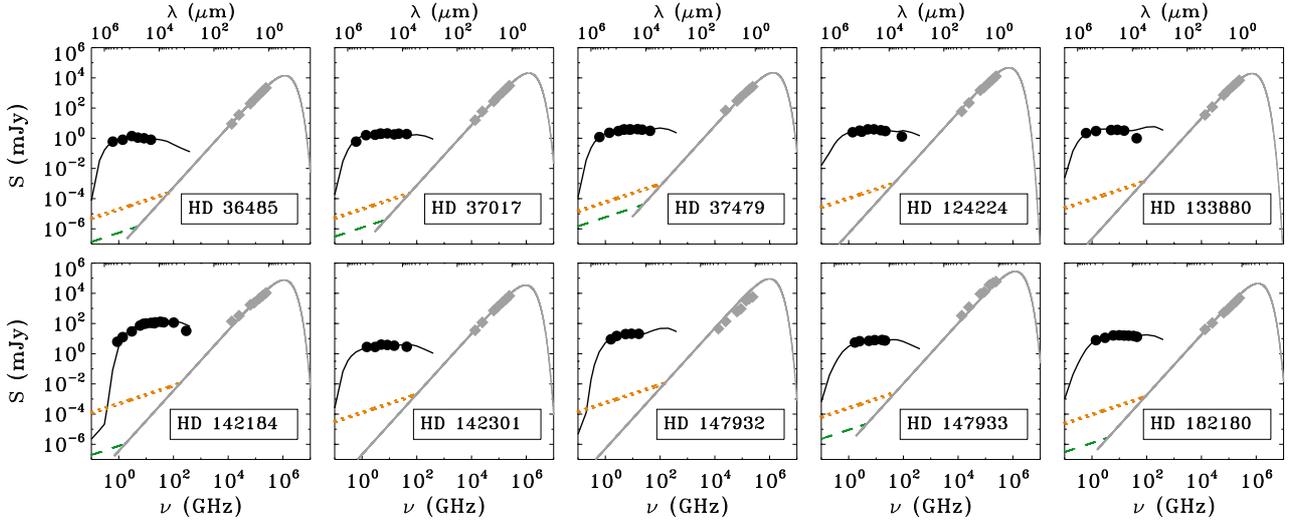}}
\caption{
{Black line: gyro-synchrotron spectra already shown in Fig.~\ref{spe10sou};
grey line: black-body spectra obtained using the stellar parameters listed in Table~\ref{tab_stars}; 
orange dotted line: wind spectra calculated using the relation given by \citet{scuderi_etal98} and 
adopting the spherical mass loss rate listed in Table~\ref{tab_stars_spectra};
green dashed line: wind spectra calculated using the theoretical wind \citep{krticka14},
the expected wind spectra of some stars are lower than the black-body emission.
Filled bullets are the radio measurements,
filled gray diamonds represent the WISE and 2MASS infrared measurements.}
}
\label{fig_sed}
\end{figure*}

\section{The role of the wind}
\label{sec_wind}

\subsection{Indirect estimation of the wind} 
\label{sec_wind_est}

{The MCWS paradigm successfully explains 
many observational features of the early-type magnetic stars,
from the radio to the X-ray regime \citep{babel_montmerle97}.
In particular, the thermal X-ray emission
from the magnetically confined wind is produced by the strong shock
occurring close to the magnetic equatorial plane \citep{ud-doula_etal14}.
Following the Rankine-Hugoniot condition,
the temperature of the X-ray radiating plasma ($T_{\mathrm X}$) is related to the velocity of the colliding wind streams
($v_{\mathrm{wind}}$)
arising from the stellar hemispheres with opposite magnetic polarity by the relation 
$T_{\mathrm X} \approx 14(v_{\mathrm{wind}} / 10^3 \,{\mathrm{km\,s^{-1}}})^2$ MK. 
Approximating $v_{\mathrm{wind}}$ with $v_{\infty}$,
the terminal wind velocity 
can be indirectly constrained by the measured temperature
of the thermal plasma emitting X-rays,
that is $T_{\mathrm X} \approx 10$ MK
\citep{
oskinova_etal11, 
pillitteri_etal16, 
pillitteri_etal17, 
leto_etal17, 
leto_etal18, 
robrade_etal18}.
The corresponding value of
the terminal wind velocity  
is $v_{\infty} \approx 900$ km\,s$^{-1}$,
that is reasonable for main sequence B type stars \citep{prinja89}.
}

In the paradigm associated with wind magnetic confinement, 
non-thermal acceleration is placed at the equatorial Alfv\'{e}n
radius. The $L$-shell parameter from our models
(Sec.~\ref{sec_spectra_sim}) can be used as an indirect estimate
of $R_{\mathrm{A}}$.  
Thus, we can quantify the spherical wind 
{of the stars analyzed in Sec.~\ref{sec_subsample} by
equating the kinetic energy density of the wind (ram pressure plus centrifugal component)
to the magnetic energy density:
$p_{\mathrm{ram}}+1/2 ( \rho \omega^2 d^2 ) = B^2/8\pi$, where $\rho$ is the wind density, 
$\omega$ is angular rotation speed, and $d$ 
the distance of a generic point located on the magnetic equatorial plane from the rotation axis.
As a consequence of the ORM geometry
the equatorial Alfv\'{e}n radius is a function of the magnetic longitude \citep{trigilio_etal04}.
Comparing the longitudinal average of $R_{\mathrm{A}}$ with the $L$-shell size
of the individual stars required to reproduce the measured radio spectra,
we indirectly derive the mass loss rate of the spherical wind ($\dot M_{\mathrm L}$)
that is able to exceed the magnetic tension at the appropriate distance to originate a radio emitting magnetosphere that reproduces the spectrum.
The values of $\dot M_{\mathrm L}$ thereby derived for the individual stars
are listed in Table~\ref{tab_stars_spectra}. The
mass loss rates are in the range $10^{-10}$--$10^{-9}$ M$_{\odot}$ yr$^{-1}$.
}

\subsection{Comparison with the theoretical wind}
\label{sec_wind_comparison}

The stars analyzed in Sec.~\ref{sec_subsample} have
effective temperatures ranging from $\approx 12$--23 kK.
{The effective temperature has a primary effect on the radiatively driven stellar wind.}
However, the radio spectra of our sample stars are well reproduced with
similar values of the magnetic shell size $L$ (Sec.~\ref{sec_spectra_sim}),
hence, no correlation seems to exist between $L$-shell and $T_{\mathrm {eff}}$.

Estimated values of {the mass loss rate ($\dot{M_{\mathrm L}}$ listed in Table~\ref{tab_stars_spectra})} 
for the spherical winds of non-magnetic
counterparts to the individual stars of our sub-sample 
are shown
in the top panel of Fig.~\ref{wsim_teo}, as a function of the corresponding
effective temperature.
{The rather limited spread in wind mass-loss
rates as estimated from the radio spectral models  
stands in
substantial contrast with values expected from radiative wind
driving, mainly owing to the rather large spread in stellar temperatures and consequently
luminosities of the sub-sample.  As an example, consider HD\,147933,
with a luminosity $L_{\ast} \approx 4000$~L$_{\odot}$
\citep{pillitteri_etal18} as opposed to HD\,133880, with $L_{\ast}
\approx 80$~L$_{\odot}$ \citep{netopil_etal17}.  Hence, HD\,147933
is about 2 orders of magnitude brighter than HD\,133880.  
According to the CAK \citep{cak75} framework of the radiatively driven wind acceleration, 
the expected mass-loss rate is related to the stellar luminosity as $\dot{M} \propto L_{\ast}^{1/\alpha} $, 
with $\alpha \approx 0.6$ \citep{puls_etal08,ud-doula_etal14}, the expected mass-loss
rate from HD\,147933 would be about 3 orders of magnitudes higher than for HD\,133880.
Further,} 
{there is a large discrepancy between the mass loss rates empirically estimated
from radio {(reported in Table~\ref{tab_stars_spectra})} and those predicted theoretically.
}
\citet{krticka14} theoretically 
derived the dependence of mass-loss rates on
$T_{\mathrm {eff}}$ for BA-type dwarfs,
{given by the relation: $\log \dot{M}=a+b T_4+cT_4^2$, with $T_4=T_{\mathrm{eff}}/(10^4\,{\mathrm K})$, 
and where
$a = -22.7$, 
$b = 8.96$, and
$c = -1.42$.}
Their predictions are in
good agreement with the empirically derived $\dot{M}$ of magnetic B-type
stars \citep{oskinova_etal11}. 
{Considering the top panel of Fig.~\ref{wsim_teo}, 
the $\dot{M}$ required to
reproduce the radio spectra} {(see Sec.~\ref{sec_wind_est})}
are significantly higher than expected, with
the gap widening at lower $T_{\mathrm {eff}}$. The discrepancy is especially
large at the lowest $T_{\mathrm {eff}}$ corresponding to Ap stars. 
For instance, for a
template A0p star, CU\,Vir (HD\,124224), \citet{krticka_etal19} derive the upper limit
$\dot{M}=10^{-12}\,M_\odot$\,yr$^{-1}$, which is orders of magnitude lower than
that required by our radio spectra models.

To further investigate the physical conditions 
in atmosphere of our sample stars,
in the bottom panel of Fig.~\ref{wsim_teo} we 
indicate the positions of stars in our sub-sample on
the $\log g / T_{\mathrm{eff}}$ diagram. The 
surface gravity ($\log g$) was retrieved from the literature for  
the individual stars \citep{bailey_etal12,grunhut_etal12,rivinius_etal13,alecian_etal14,kochukhov_etal14,
pillitteri_etal18,shultz_etal19_485}. The values of
$\dot{M}$ for main sequence B-type stars has been theoretically
predicted by \citet{babel96},
who recognized three wind regimes, 
namely: a homogeneous wind; a chemically inhomogeneous wind; and static 
atmosphere. 
These three wind regimes depend on the combined effects 
of temperature and surface gravity.
Based on results  from 
Fig.~1 of \citet{hunger_groote99}, the qualitative effect of chemical
anomalies is roughy to shift down the lower boundary curve taken
from Fig.~6 of \citet{babel96}.
In the bottom panel of Fig.~\ref{wsim_teo}
the down-shifted curves are pictured
using the light grey with a gradually decreasing intensity.

The location on  the $\log g / T_{\mathrm{eff}}$ diagram of 
the ten stars from our sub-sample is in accordance with the
mass loss recipe of \citet{krticka14}.
Note that about $85\%$ of the stars in our complete
sample have $T_{\mathrm {eff}} \lessapprox 20$ kK, 
in which
case only weak metallic {($\dot M \lessapprox 10^{-14}$ M$_{\odot}$\,yr$^{-1}$)} winds are expected \citep{babel96}.
Accounting for the corresponding magnetic field strength,
we expect Alfv\'{e}n radii larger than 
the average value retrieved by {calculating the synthetic} 
spectra 
($R_{\mathrm A} \approx 14$ stellar radii).

In late B and early A stars, the inhomogeneous 
(hydrogen-free) weak wind \citep{babel95,babel96} would continuously 
deposit a small amount of metal ions into their centrifugally supported 
magnetospheres.
The secular accumulation of trapped ionized material 
eventually fills the magnetosphere. 
The density distribution 
of the magnetically confined circumstellar plasma around hot 
magnetic stars has been theoretically studied by
\citet{preuss_etal04} and \citet{townsend_owocki05}.
The thermal plasma 
accumulated by the wind at low magnetic latitudes 
might be the cause 
of significant departures from the simple dipole topologies.
Far from the star, the magnetic tension might no longer be able to 
confine this thermal material, leading to centrifugal breakout 
\citep{ud-doula_etal06}, with consequent local magnetic reconnection 
that is likely a source of plasma acceleration. 
\citet{shultz_etal20} and \citet{owocki_etal20} have demonstrated that the onset, 
emission strength scaling, and line profile morphologies of H$\alpha$ emission 
from centrifugal magnetospheres can only be explained 
if mass-balancing is achieved by continuous centrifugal breakout events 
occurring at small spatial scales throughout the centrifugal magnetosphere.

\subsection{The wind spectra and the stellar SEDs}
\label{sec_wind_sed}

{To further investigate the wind mass loss from early-type magnetic stars, we have taken
into account possible effects due to the ionized material carried out by the radiatively driven stellar wind.
We compared the wind thermal emission 
with the corresponding non-thermal radio emission.
Both wind regimes discussed in Sec.~\ref{sec_wind_comparison} have been analyzed.
Further, we extended to the radio regime the expected emission from the stellar surface
modeled by a simple black-body spectrum. 

Radio emission from stellar winds was theoretically studied under simplified assumptions.
The radio flux of a spherical wind increases as a function of the frequency like $S_{\nu} \propto \nu^{0.6}$  \citep{wright_barlow75,panagia_felli75}.
In Appendix~\ref{wind_appendix} the 
equation of the wind spectrum is explicitly reported (Eq.~\ref{eq_wind_spectrum}).
The wind spectra calculated using Eq.~\ref{eq_wind_spectrum} and adopting the indirect estimation of the spherical wind
of the individual stars ($\dot M _{L}$ listed in Table~\ref{tab_stars_spectra}) are shown  in
Fig.~\ref{fig_sed},
pictured by the orange dotted lines.  
Looking at the figure, it is clear that the expected wind radio emission 
from the B/A magnetic stars analyzed here is faint.
The wind spectra
cover almost the same spectral range of the non-thermal radio emission,
but are orders of magnitudes less bright.

For the ten stars analyzed in Sec.~\ref{sec_subsample},
the spectral energy distributions (SED), from the radio to the UV domain, 
have been calculated combining the radio spectra (Sec.~\ref{sec_spectra_sim})
and the blackbody spectra ($T_{\mathrm{eff}}$ listed in Table~\ref{tab_stars})
radiated from the surface scaled at the stellar distance.
In the figure, the radio and infrared measurements 
have also been reported. To minimize 
for possible absorption effects, interstellar or intrinsic,
we collected near- and mid-infrared measurements, from the
Two Micron All Sky Survey     
(2MASS)
\citep{2mass} and
the Wide-field Infrared Survey Explorer 
(WISE) 
\citep{allwise} catalogues;
these measurements are shown in Fig.~\ref{fig_sed}. 

The wind spectra corresponding to the theoretical radiative wind mass-loss rates, 
expected on the basis of the corresponding stellar temperatures,
have been calculated too.
As expected by the low level of the theoretical radiative wind from 
B/A spectral type stars,
the wind spectra are much fainter (green dashed lines shown Fig.~\ref{fig_sed}).
In some cases the wind mass loss rate is so low as to radiate
less than the black-body radiation of the stellar surface, at the radio spectral range analyzed.
In these cases the corresponding wind spectra have not been shown.

\begin{figure}
\resizebox{\hsize}{!}{\includegraphics{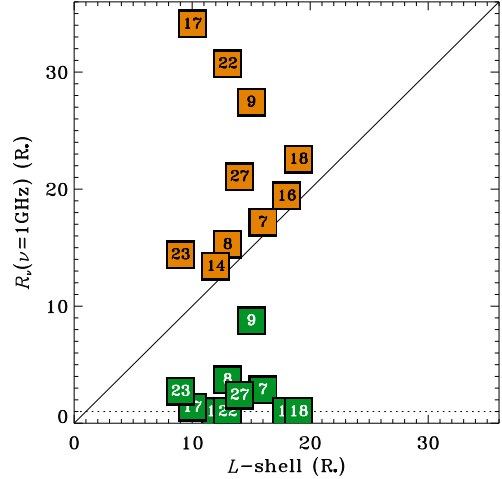}}
\caption{{Radii of the radio photosphere calculated at the representative frequency $\nu=1$ GHz as a function of the 
corresponding magnetospheric size ($L$-shell)
required to reproduce the radio spectra of the non-thermal radio emission.
The $R_{\nu}$ values have been derived in two cases: assuming the wind mass loss rate 
theoretically expected (green symbols); and assuming the $\dot M_L$ required to break the stellar magnetic field lines
at the distance equal to the $L$-shell (orange symbols).
The dotted line locates the lower limit of $R_{\nu}$, coinciding with the stellar radius.
The solid line separates the case of a wind radio photosphere that incorporates
the region radiating non-thermal radio emission
(upper left) from
the case where the wind emitting region is located inside (bottom right).
}}
\label{fig_r_vs_l}
\end{figure}

The early type magnetic stars are surrounded by 
large scale dipole-like magnetic field, in this case
the spherical wind assumption is a rough schematization of the real case.
In fact, the wind material can be lost from the polar caps only,
where the magnetic field lines are open.
At the lower magnetic latitudes, the wind is trapped and channeled by the closed field lines. 
The radiative wind in presence of magnetic field produces the typical observable features, 
from the X-ray to the radio, successfully explained by the MCWS model.

The effective wind mass loss rate of the early-type magnetic stars
is expected lower than a simple spherical wind.
Therefore, it is plausible to expect that the spherical wind assumption 
produce overestimated effects. 
In any case, also adopting the simplified spherical assumption,
the wind emission is never comparable to the emission level of the non-thermal
radio emission produced by the MCWS model. 
The ionized matter released by the polar caps 
might produce frequency-dependent absorption effects 
for the non-thermal radio emission.
The theory of spherical stellar winds predicts that
each radio frequency of
the wind spectrum arises from a well-constrained emitting region,
similarly to a radio photosphere where the optical depth is $\tau \approx 0.4$ \citep{panagia_felli75}.
The radio photosphere location is a function of the observing frequency,
with the radius of this emitting region decreasing as the frequency increases,
in accordance with the relation
$R_{\nu} \propto \nu^{-0.7}$. Adopting the spherical wind simplified assumption,
the dependence of the size of the wind-emitting region on the mass-loss rate and on other stellar parameters
is explicitly reported in Eq.~\ref{eq_wind_radius}.
The high frequency cutoff of the wind spectrum is defined by the frequency radiated 
from the wind region that meets the stellar surface.

To account for possible absorption effects, we calculated the size of
the wind radio photosphere at $\nu=1$ GHz. We chose this frequency because the bulk
of the non-thermal radio emission is produced at $\nu >1$ GHz.
The radii of the radio photosphere as a function of the size of the 
magnetosphere ($L$-shell parameter) where the non-thermal radio emission arises
are shown in Fig.~\ref{fig_r_vs_l}.
The $R_{\mathrm{\nu=1\,GHz}}$ values 
have been calculated for each star for both wind regimes:
the $\dot M$ theoretically expected from the stellar temperatures;
and the $\dot M_L$ indirectly derived from the non-thermal spectra (listed in Table~\ref{tab_stars_spectra}).
Looking at Fig.~\ref{fig_r_vs_l} it is evident that the radio photosphere calculated using $\dot M$
is contained within the non thermal emitting region,
making any possible absorption effect negligible.

The non-thermal radio spectra calculations have been performed under the assumption that,
 outside the middle-magnetosphere,
the absorption effects of thermal plasma can be neglected. 
This corresponds to assuming the electromagnetic waves propagate in vacuum.
The accordance between the observed and the calculated spectra
support this assumption.
It follows that the negligible absorption effect predicted by the weak wind regime
is in accordance with our model assumption.
This is a further evidence that the 
mass loss rate values indirectly derived from the non-thermal radio emission ($\dot M_L$) are unreliable.
}

\section{Is radio emission age-dependent?}
\label{sec_age}

{As discussed in Sec.~\ref{sec_wind},
the over-simplified assumption that the size of the radio emitting magnetosphere
coincides with the Alfv\'{e}n radius has something wrong.
If the wind is the plasma source which continuously accumulates within the inner magnetosphere,
then, non-thermal}
radio emission from early-type magnetic stars 
might be potentially related to the stellar age. As suggested 
in Sec.~\ref{sec_wind_comparison}, the wind plasma accumulation produces centrifugal breakout 
events \citep{shultz_etal20,owocki_etal20}
that might make the corresponding stellar magnetosphere radio-loud.

\begin{figure}
\resizebox{\hsize}{!}{\includegraphics{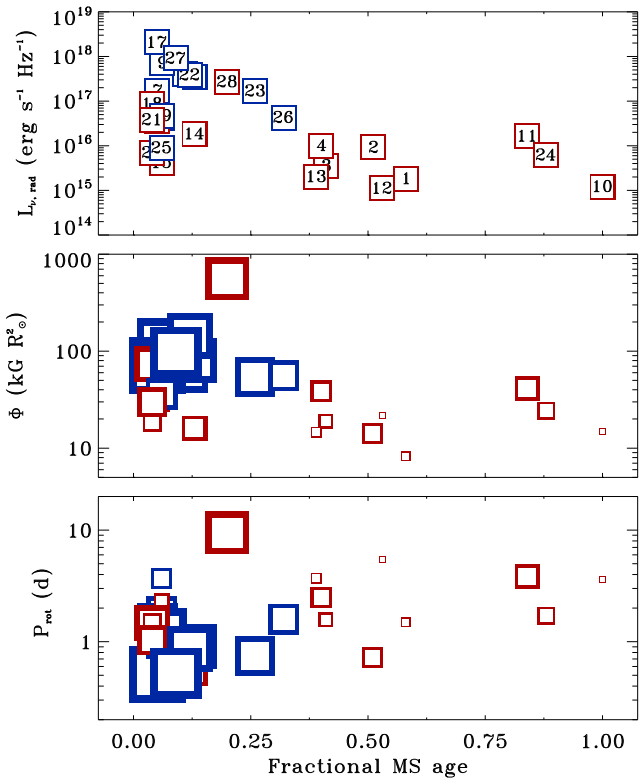}}
\caption{Top panel: Radio luminosities as a function of the fractional main sequence age,
where blue symbols refer to stars earlier than B8 and red symbols to the cooler stars.
Middle and bottom panels show magnetic fluxes the rotation periods as a function of 
fractional age. The size of the symbol is proportional to the radio 
luminosity of the individual star.}
\label{fig_tau}
\end{figure}

To search for a possible age dependence, the measured radio 
luminosities are shown in the top panel of Fig.~\ref{fig_tau} as a 
function of the corresponding fractional main sequence age (as 
listed in Table~\ref{tab_stars}). 
The stars earlier than spectral type B8 have 
been marked using blue symbols, while red symbols have been 
used for cooler stars. Looking at the top panel of this figure, 
no significant relation appears evident between the stellar radio 
luminosity and corresponding stellar age for the hotter stars,
whereas a possible decaying trend is suggested for the cooler stars.
This effect resembles the results obtained by \citet{fossati_etal16},
moreover this closely matches to the behavior of the H$\alpha$ emission 
discovered by \citet{shultz_etal20}.
Studying a large sample of massive and hot main sequence magnetic 
stars, \citet{fossati_etal16} found that large-scale magnetic fields suffered 
a time-dependent decaying trend,
effect that was further confirmed by \citet{shultz_etal19_490}.
In particular, the magnetic flux $\Phi$ ($=B_{\mathrm 
p} R_{\ast}^2$) decreases more steeply than expected from the 
magnetic flux conservation law, as predicted by the expansion of the stellar radius along the MS evolution.
Further, \citet{sikora_etal19a} indicated 
that the chemical surface composition of early type magnetic 
stars is strictly related with the stellar age, particularly for the cooler stars.

\begin{figure*}
\resizebox{\hsize}{!}{\includegraphics{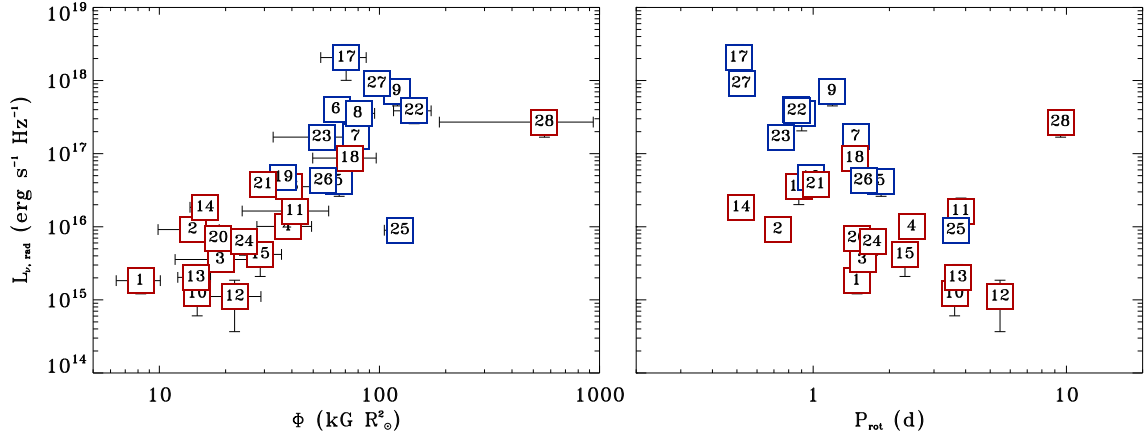}}
\caption{Left panel: the radio luminosity as a function of the 
magnetic flux ($\Phi$). Right panel: radio luminosity as a function 
of the rotation period ($P_{\mathrm{rot}}$).
The blue symbols refer to the hot stars, and the red to the cooler stars (SP$>$B8).}
\label{lmp}
\end{figure*}

Although the sample of early-type magnetic stars analyzed here
is small, 
we have searched for possible age-dependent effects on 
the magnetic flux and the stellar rotation.
As already well studied in this class of stars,
the stellar rotation period is expected to increase dramatically over time
\citep{sikora_etal19b,shultz_etal19_490}.
In the middle and 
bottom panels of Fig.~\ref{fig_tau}, we show $\Phi$ and $P_{\mathrm{rot}}$ 
as a function of the fractional main sequence stellar age. The 
corresponding stellar radio luminosity is indicated by the symbol size.
For the hotter stars, the two
analyzed stellar parameters do not show any significant age-dependent
effects, whereas the magnetic field survives on the colder
stars.
Among the small sample of stars here analyzed, 
at young ages all magnetic field strengths are possible,
but at old ages, there are few stars with strong fields, 
in particular,  
the older stars in our sample are also cooler stars (see Fig.~\ref{fig_tau}).
The same trends seem to be 
confirmed by the radio. In fact, there is a dearth of both old and 
strong radio emitters.

The large-scale magnetic fields of the early-type magnetic stars 
might significantly affect the stellar evolution.
\citet{shultz_etal19_490} performed an extensive study on the dependences 
of the stellar magnetic flux as a function of various stellar parameters. 
They conclude that massive stars show a more 
pronounced magnetic flux decreasing with the age.
Further, magnetic braking effects significantly 
reduce the stellar rotation period during the star's life.

Although the small number of magnetic stars 
known as radio sources may introduce a selection bias, and a
large fraction of stars in our sample have similar ages, their 
range of radio luminosities are fairly broad (see top panel of Fig.~\ref{fig_tau}). 
This suggests that any
connection between radio emission, magnetic field strength, rotation 
period, and age is not simple.
The only relevant effect which emerged from the analysis of our small sample
is the possible rough dependence of the radio luminosity 
with the magnetic flux and rotation period shown by the stars having 
similar ages. In particular, stars with a short rotation period and 
a high magnetic flux produce
stronger radio emission
(see middle and bottom panels of Fig.~\ref{fig_tau}).

\section{The effect of the rotating dipole}
\label{sec_rotation}

As discussed in
Sec.~\ref{sec_wind}, the wind 
is not the lead actor in the production of 
non-thermal radio emission. By contrast there may be a possible 
dependence on other stellar parameters that mainly characterize the 
rotation of the dipole-like magnetic field (Sec.~\ref{sec_age}).

The sample analyzed in this paper has heterogenous parameters. 
To well characterize the magnetic properties of the sample, 
a reliable parameter is the magnetic flux ($\Phi=B_{\mathrm 
p} R_{\ast}^2$), which combines  
both the effects due to the 
magnetic field strength and the stellar size. In the two panels 
of Fig.~\ref{lmp}, the stellar radio luminosities are displayed as 
a function of $\Phi$ (left panel) and $P_{\mathrm{rot}}$ (right 
panel). 
The obvious trends seen in Fig.~\ref{lmp}, confirm the dependence
of radio emission on these two parameters. 
In particular, $L_{\nu,\mathrm{rad}}$ has a positive 
dependence with $\Phi$ and a negative one with $P_{\mathrm{rot}}$. 
These trends suggest identifying a parametrization that incorporates 
both these two parameters. The obvious way is to take their ratio.

\begin{figure}
\resizebox{\hsize}{!}{\includegraphics{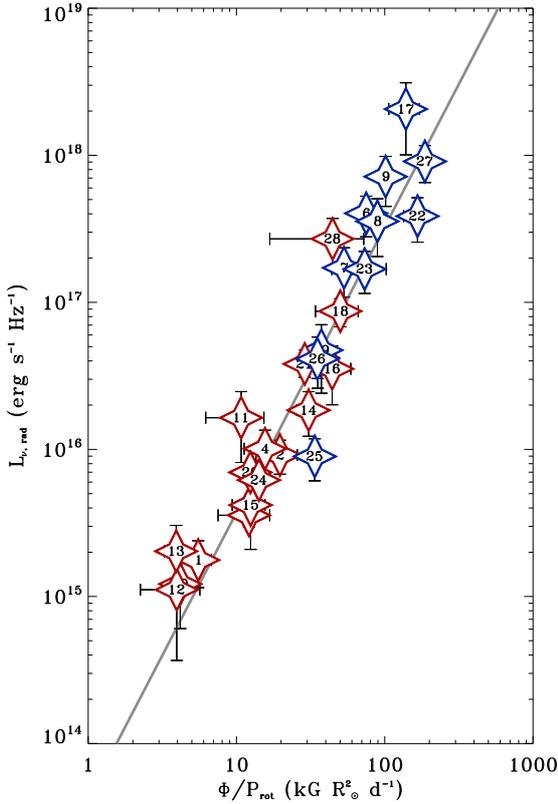}}
\caption{
Radio luminosity of the early-type magnetic stars as a function of 
the ratio $\Phi/P_{\mathrm{rot}}$. The solid grey line is the 
power-law that best fits the two parameters. As in Figs.~\ref{fig_tau} 
and ~\ref{lmp}, the blue symbols refer to the hot stars, while 
the red symbols refer to the cooler stars (SP$>$B8).}
\label{lum_mv_mcp}
\end{figure}

In Fig.~\ref{lum_mv_mcp} the radio luminosities of the magnetic 
stars are reported as a function of what we are calling the ``magnetic flux rate'',
${\Phi} / {P_{\mathrm {rot}}}$ (kG~R$_{\odot}^2$~d$^{-1}$).
Already visual inspection of Fig.~\ref{lum_mv_mcp} reveals a strong 
correlation. To quantify the correlation,  we calculated  the Pearson 
correlation coefficient ($r$) and the Spearman's rank correlation 
coefficient ($r_{\mathrm s}$). The $r$ statistical parameter is only
sensitive to a linear relationship,  whereas $r_{\mathrm s}$ 
is sensitive to any possible monotonic relation between the two 
analyzed variables. These two statistical parameters range between 
$-1$ and 1, with absolute values $\approx 0.7$, or higher, 
indicating a high correlation. 

The correlation parameters indicated that the radio luminosity is 
strongly correlated to the ratio of magnetic flux to 
rotation period. In particular, $r \approx 0.75$ and $r_{\mathrm 
s} \approx 1$, suggesting that a non-linear relation exists.
The scaling relation that  best fits the data is:
\begin{equation}
L_{\nu,\mathrm{rad}} = 10 ^{\alpha}  
\left(\frac{B_{\mathrm p}}{{\mathrm{kG}}}\right)^{\beta} 
\left(\frac{R_{\ast}} {{\mathrm{R_{\odot}}}}\right)^{2\beta} 
\left(\frac{P_{\mathrm {rot}}} {\mathrm d} \right)^{-\beta}
(\mathrm{erg} ~ \mathrm{s}^{-1} ~ \mathrm{Hz}^{-1})
\label{eq_rel_scal}
\end{equation}
\noindent with $\alpha=13.6 \pm 0.1$ and $\beta=1.94 \pm 0.07$.

It is worth noting that following the Faraday-Lenz law, the empirical 
parameter $\Phi/P_{\mathrm {rot}}$ has the physical dimension of 
an electromotive force (e.m.f.$=-d{\Phi}/d{t}$). Even if in a co-rotating 
reference frame the stellar magnetic flux is not expected to be 
variable over time: a spinning dipolar magnetosphere has close 
similarities with Faraday disk dynamo, where the 
rotation of a conductive disk within a stable magnetic field, having 
the magnetic vector orientation aligned with the disk rotation axis, 
produces an e.m.f. (see experiments of \citealp{muller14}). We analyze 
the rotating magnetospheres from a reverse point-of-view.  In a 
reference frame anchored to a free test charge located in the 
equatorial plane, with distance $R$ from the rotation axis,
the magnetic field lines will be seen moving at velocity $v= \omega 
R$ ($\omega=2 \pi / P_{\mathrm{rot}}$ rotational angular frequency).
Assuming the ideal case of aligned dipole and spin axes, the magnetic and equatorial planes are coinciding.
Then in the magnetic equatorial plane the magnetic field vectors are orthogonal to the plane and also
to the velocity vector.
Assuming the simplified homogeneous magnetic field condition,
it follows that the test charge located at the stellar equator 
embedded within a
homogeneous field (strength $B_{\mathrm {eq}}$) will be 
subjected to an induced electromotive force \citep{jackson62}
${\mathrm{e.m.f.}}=-d{\Phi}/d{t}=\oint {\bf v} \times {\bf B}\, {d}{\bf l}$,
that under the above simplified conditions can be written as:
${\mathrm{e.m.f.}}=\int_0^{R_{\ast}} v\, B_{\mathrm{eq}}\, {d}r = \int_0^{R_{\ast}}  \omega\, r\, B_{\mathrm{eq}}\, {d}r  = 1/2 \omega B_{\mathrm{eq}} R_{\ast}^2$ ($\propto \Phi / P_{\mathrm{rot}}$),
(for details see \citealp{muller14}).
{Note that for a real
magnetized plasma, the free charged particles cannot cross the magnetic field lines.
The spinning stellar dipole plunged in the magnetospheric plasma can be treated
following an ideal magnetohydrodynamic approach.
The above discussion is only used 
to illustrate that a spinning dipole may be a source of
electric voltage, potentially able to originate electric currents. In fact,}
in astrophysical applications, the electrodynamics of spinning dipoles 
have been the subject of many theoretical studies 
\citep{goldreich_lynden-bell69,goldreich_julian69,ruderman_sutherland75}.

{The rough schematization performed above may be useful to perform
order of magnitude relation between astrophysical observables (i.e. the radio luminosity)
and general stellar parameters (i.e. the the stellar rotation period, the radius, and the magnetic field strength).
In fact, the relation
$\omega B_{\mathrm {eq}} R^2_{\ast}$, that resemble an electromotive force,}
is a useful scaling parametrization 
for the realistic electrodynamics of spinning magnetospheres.
The stars analyzed in this paper typically have a dipolar 
magnetic field topology, with $B_{\mathrm {eq}}=B_{\mathrm P}/2$. 
Thus, the empirical  parameter $\Phi/P_{\mathrm {rot}}$ can be 
easily transformed to the physical parameter $\omega B_{\mathrm{eq}} 
R_{\ast}^2$. If the rotation period is given in seconds, the 
equatorial magnetic field strength in Tesla, and the stellar radius 
in meters, the parametric electromotive force (named e.m.f.$^{\ast}$) 
has the physical unit of Volts. In Fig.~\ref{lum_mv}, the stellar radio 
luminosities are now reported as a function of the corresponding 
values of e.m.f.$^{\ast}$ and given in unit of $10^6$~V. The 
e.m.f.$^*$ is not the true electromotive force acting within the stellar magnetosphere 
(its study is out of the scope of this paper)
but this is rather
a useful parameter with which to quantify the capability of rotating magnetospheres 
for producing electric currents.

\section{Other ordered magnetospheres}
\label{other_cases}

To test whether the result reported in Sec.~\ref{sec_rotation} holds
only for the early-type magnetic stars or if it is also valid in a wider
context, we analyzed the incoherent radio emission from other types
of stars (and one planet) characterized by large-scale and well-ordered
magnetospheres, that are mainly described by a simple dipole magnetic
field topology.

\subsection{The Case of Jupiter}
\label{jupiter}

The giant planet Jupiter is surrounded by a large-scale (almost dipolar) magnetic 
field, with the dipole axis nearly aligned with the planetary rotation 
axis (tilt angle $\approx 10^{\circ}$). Compared to the hot magnetic 
stars, Jupiter has a weak magnetic field, with $B_{\mathrm{J}}\approx 
4.17$ G at the equator \citep{connerney_etal18}, corresponding to 
$B_{\mathrm P} \approx 0.008$ kG. 
Jupiter is surrounded by ionized matter that comes from the volcanic 
moon Io (Io plasma torus) or captured from the Solar wind,
which is affected by the planetary magnetic field out to a large distance (tens of Jupiter radii).

The jovian magnetosphere is a site of plasma phenomena responsible 
for Jupiter's radio emission, which has been well-studied using 
ground-based observing facilities and radiometers onboard spacecrafts.
Its radio spectrum is dominated at the low-frequency side by 
coherent auroral radio emission \citep{zarka04}, and at high 
frequencies ($\nu \gtrapprox 10$ GHz) by the black body thermal 
emission of Jupiter's atmosphere. At intermediate frequencies, the 
incoherent synchrotron emission is another component of 
Jupiter's radio spectrum. The multi-wavelength VLA images of the 
planet Jupiter show the spatial distribution of the non-thermal 
radio emission component from Jupiter \citep{de_pater91}. This 
non-thermal component originates from the inner radiation belt 
located at $L < 5$ Jupiter radii, inside the orbit of the moon Io.

The non-thermal radio component was disentangled from the thermal one and
the wide band non-thermal jovian radio spectrum was discussed by \citet{de_pater04}. We note that the 
spectral shape qualitatively resembles the overall behavior of the 
radio spectra of early-type magnetic stars (see Fig.~\ref{spe10sou}), but is tuned at lower 
frequencies. The reference frequency of the
Jupiter's incoherent non-thermal emission is in fact $\approx 1.6$ 
GHz. The average luminosity of the incoherent radio emission is 
$L_{\nu,{\mathrm{rad}}}=1.8 (0.4)\times 10^{6}$ erg\,s$^{-1}$\,Hz$^{-1}$,
obtained from scaling the multifrequency fluxes measurements to the 
distance from Earth of 4.04 a.u \citep{de_pater_etal03,de_pater_dunn03}.

Jupiter's equatorial radius and magnetosphere rotation period are
$R_{\mathrm{J}}=71492$ km and $P_{\mathrm{J}}=9$\,h 55\,min 29\,s.
In MKS unit, the effective electromotive force (as defined in 
Sec.~\ref{sec_rotation}) produced by the rotation of the Jupiter's 
magnetosphere is ${\mathrm{e.m.f.}^{\ast}}=376$ MV. This coincides 
with the magnetic potential energy available from the jovian 
magnetosphere's corotation, defined by \citet{khurana_etal04} as 
$\omega_{\mathrm{J}} B_{\mathrm{J}} R_{\mathrm{J}}^2$ (with 
$\omega_{\mathrm{J}}=2 \pi / P_{\mathrm{J}}$ as Jupiter's rotational 
angular frequency). Further, the rotation of the jovian magnetosphere 
has been recognized as a source of stationary field-aligned current 
systems that drive the main auroral oval centered on the magnetic 
poles \citep{hill01,cowley_bunce01}.

The radio luminosity of Jupiter's non-thermal incoherent radio emission is 
reported in Fig.~\ref{lum_mv} as a function of the corresponding 
${\mathrm{e.m.f.}^{\ast}}$. Interestingly in the $L_{\nu,{\mathrm{rad}}} 
/ {\mathrm{e.m.f.}^{\ast}}$ diagram, the planet Jupiter is situated 
within the uncertainty range of the scaling relation retrieved by 
the study of the early-type magnetic stars extrapolated down to the 
low radio luminosity level of Jupiter.

\begin{table*}
\caption[ ]{Sample of cool dwarfs and corresponding parameters.}
\label{tab_ucd}
\footnotesize
\begin{tabular}{l c c c l l l c}
\hline
      ~                                      &                                       &                &$D$               &$R_{\ast}$                &$P_{\mathrm{rot}}^{\dag}$   &$B_{\mathrm p}$        &$L_{\nu,\mathrm{rad}}$       \\
        2MASS                          &Common or alt. name    & Sp. type  &(pc)               &(R$_{\odot}$)             &(hrs)                          & (kG)                             &(erg s$^{-1}$ Hz$^{-1}$)     \\
\hline 
       J00242463-0158201     & BRI\,0021-02            &M9.5V         &$12.50(0.03)$          &$0.113(0.005)^{(1)}$     &$\approx 5^{(4)}$              &$\gtrapprox 3$                    & $1.6(0.3)\times 10^{13}$    \\                   
         J00361617+1821104     & LSPM J0036+1821            &L3.5         &~~$8.74(0.02)$          &$0.118(0.007)^{(1)}$     &$3.0(7)^{(4)}$                    &$\gtrapprox 3$                    & $2.4(0.2)\times 10^{13}$    \\
     J10481463-3956062       &DENIS\,J1048.0-3956   &M8.5Ve       &~~~~~~$4.045(0.002)$    &$0.096(0.004)^{(1)}$     &$\lessapprox 5^{\dag\dag}$        &$\gtrapprox 3$                    & ~~~~~~$4(2)\times 10^{12}$    \\
       J15010818+2250020      &TVLM\,513-46        &M8.5V         &$10.70(0.02)$      &$0.103(0.006)^{(2)}$     &$1.959574(2)^{(5)}$             &$\gtrapprox 3$                    & ~~~~~~$3(1)\times 10^{13}$    \\
       J18353790+3259545     & LSR\,J1835+3259         &M8.5V         &$~~~~~~5.687(0.003)$  &$0.096(0.004)^{(1)}$     &$2.845(3)^{(4)}$                &$\approx 5^{(7)}$            & $2.03(0.03)\times 10^{13}$ ~~~ \\
       J22011310+2818248     & V374\,Peg        &M3.5V         &~~~~~~$9.104(0.005)$          &$0.34(0.01)^{(3)}$     &$10.69570(5)^{(6)}$              &$\approx 3^{(8)}$            & $~~~~~~7(3)\times 10^{13}$ \\
\hline
\end{tabular}
\begin{list}{}{}
\item[Notes:] 
$^{\dag}$ The uncertainty of the rotation  periods are referred to the last digit.
$^{\dag\dag}$ Upper limit of the rotation period estimated by the measured projected rotational velocity, $v \sin i=25 \pm 5$ km s$^{-1}$ \citep{stelzer_etal12}.
\item[References:]
$^{(1)}$\citealp{cifuentes_eta20}; 
$^{(2)}$\citealp{hallinan_etal08}; 
$^{(3)}$\citealp{morin_etal08a}; 
$^{(4)}$\citealp{harding_etal13}; 
$^{(5)}$\citealp{wolszczan_route14}; 
$^{(6)}$\citealp{morin_etal08b}; 
$^{(7)}$\citealp{kuzmychov_etal17}; 
$^{(8)}$\citealp{donati06}.
\end{list}

\end{table*}

\subsection{The Case of the Ultra Cool Dwarfs}
\label{ucd}

At the bottom of the main sequence, the fully convective Ultra Cool 
stars and brown Dwarfs (UCDs) show radio bursts ascribed 
to coherent auroral emission that closely resemble the case 
of Jupiter \citep{nichols_etal12}. In some cases, the 
quiescent incoherent non-thermal radio emission component was also 
detected (\citealp{metodieva_etal17}, and references therein). The UCDs largely violate 
\citep*{williams_etal14,lynch_etal16} the Guedel-Benz relation 
\citep{guedel_benz93}, that couples the radio and X-ray luminosities 
for a large fraction of the main sequence stars (from  the F to the 
early M-type stars) characterized by Solar-like magnetic activity,
making the radio behavior of these very cold objects more similar 
to those of the magnetic stars located at the top of the main sequence
or to that of giant planets.

The UCDs are magnetic stars surrounded by large scale dipole 
dominated magnetospheres, that, like the case of Jupiter, are 
strongly affected by rotation \citep{schrijver09}. The 
measurement of an axisymmetric magnetic field was first reported 
for the M3.5 star V374\,Peg \citep{donati06}. Well-ordered, 
large-scale magnetic fields were also measured in other fully 
convective UCDs \citep{reiners_etal07,morin_etal10}.
This supports the theoretical prediction of a magnetic field generated 
by a process similar to the geodynamo operating in Jupiter and 
Earth \citep*{christensen_etal09}.

\begin{figure}
\resizebox{\hsize}{!}{\includegraphics{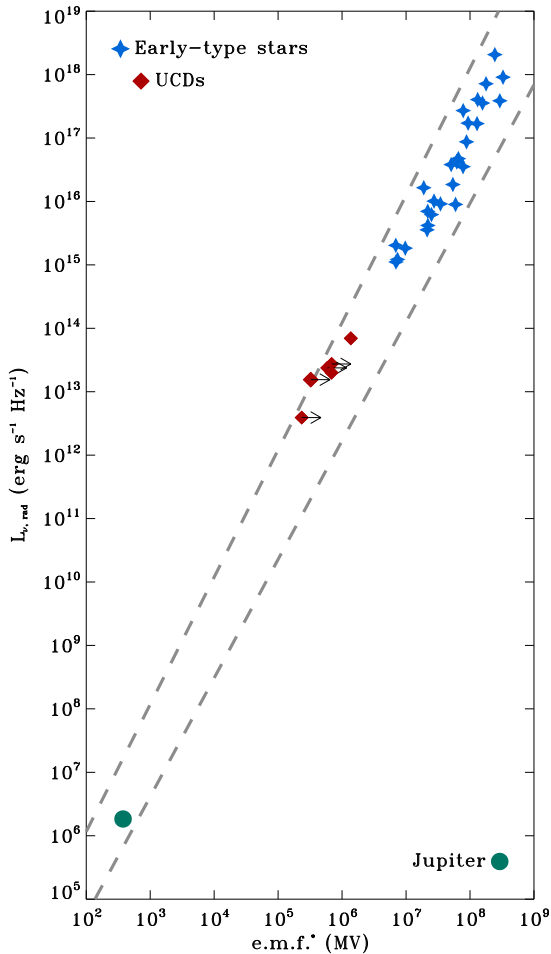}}
\caption{$L_{\nu,{\mathrm {rad}}}/{\mathrm{e.\,f.\,m.}^{\ast}}$ diagram. 
The dashed gray lines locate the boundary of the uncertainty range of the scaling relation derived by using the 
parameters of the early type magnetic stars only.}
\label{lum_mv}
\end{figure}

The UCDs are faint objects in the visual band. Further, those that 
are radio loud are also fast rotators (rotation periods of order 
an hour), allowing direct magnetic field 
measurements in only few cases. 
In some cases the magnetic fields of the UCDs can 
be indirectly constrained. In fact, following the basic theory of 
the ECM mechanism, the emission 
frequency ($\nu$) of the coherent pulses occurs close to
local gyro-frequency. Since the gyro-frequency is related to the 
local magnetic field by the relation $\nu_{\mathrm B}=2.8 \times 
B/{\mathrm {kG}}$ GHz, the polar field strength of the UCDs where 
the ECM is detected will be $B_{\mathrm p} \gtrapprox \nu/2.8$ kG.
Hence, the lower limits of the polar field strengths reported in 
Table~\ref{tab_ucd} were estimated from the maximum frequencies at 
which the coherent emission produced by the ECM was detected.
In particular, the four listed UCDs showed coherent pulses at $\nu \approx 8.4$ GHz
\citep{berger02,burgasser_putman05,hallinan_etal07,hallinan_etal08}, 
thus $B_{\mathrm p} \gtrapprox 3$ kG.

To determine whether the radio behavior of UCDs matches the scaling 
relation given by Eq.~\ref{eq_rel_scal}  
requires both reliable radio 
luminosities for the incoherent non-thermal radio emission and 
knowledge of stellar parameters, in particular the magnetic 
field strength, rotation period, and radius. 
For our analysis, we excluded UCDs that are members 
of binary systems. The incoherent fluxes reported in 
Appendix~\ref{ucd_appendix} have been used to estimate the radio 
luminosities of the six UCDs in Table~\ref{tab_ucd}. The few 
available radio measurements are mainly performed at a single 
frequency, particularly $\nu \approx 5$ or 8.4 GHz (see Appendix~\ref{ucd_appendix}).

The presence of stable electric fields within the large scale magnetospheres
of the UCDs was proposed to have a possible role on the non-thermal electron production
responsible for the radio emission (coherent and incoherent)
from these objects \citep{doyle_etal10}.
In Fig.~\ref{lum_mv}, the radio luminosities of the non-thermal 
incoherent emission of the small sample of UCDs analyzed here are 
reported as a function of their effective electromotive forces (or 
lower limits). Despite the limitations explained above, the UCD 
parameters lie in the $L_{\nu,{\mathrm 
{rad}}}/{\mathrm{e.\,f.\,m.}^{\ast}}$ diagram either within (or very 
close to) the uncertainty range extrapolated from the scaling relation 
established by the sample of early-type magnetic stars.

\section{The scaling relationship for the radio emission applied to hot Jupiters}
\label{hot_jupiters}

{Hot Jupiters are giant gaseous planets orbiting very close to their parent stars.
Due to the proximity to the star,
hot Jupiters have suitable conditions to develop strong 
star planet magnetic interactions (SPMI),
extending Bode's law,  that holds for the magnetized  planets of the solar system, to the exo-planets \citep{zarka07}.
Bode's law is an empirical scaling relation between the power of the coherent auroral radio emission from
a magnetized planet and
the power dissipated when interacts with 
the magnetized plasma released by the parent star, and vice-versa.
Due to their small orbits,
the extrapolated Bode's law predicts detectable coherent auroral radio emission from magnetized hot Jupiters
\citep{zarka07}.

The auroral radio emission is tuned at the radio frequency typical of the ECM mechanism,
that, as previously discussed (see Sec.~\ref{ucd}),
depends by the magnetic field strength of the body where it originates (star or planet).
The estimated magnetic field strength of the exo-planets belonging to the hot Jupiter class 
ranges from few G (like the case of Jupiter) up to few hundred G
\citep{yadav_etal17}. Then, the expected ECM emission frequency lies in low frequency
spectral domain (tens to a few hundred MHz).
The tentative detection by the LOFAR interferometer of highly polarized low frequency radio emission
from the hot Jupiter $\tau$\,Bootes\,b
has been recently reported \citep{turner_etal21}.
This result supports the Bode's law extension beyond the Solar system,
and further provides indirect evidence of magnetism occurring in exo-planets.

\begin{table}
\caption[ ]{Hot Jupiters parameters.} 
\label{tab_hot_jup}
\footnotesize
\begin{tabular}{@{}l c c c c@{}}
\hline
      ~                                         &$D$                          &$P_{\mathrm{rot}}$          &$B^{(5)}$                        &$R^{(5)}$                 \\
        Name                               &(pc)                          &(d)                                       & (G)                      &(R$_{\mathrm J}$)                                \\
\hline 
HD\,179949\,b                         &$27.48(0.06)^{(1)}$          &$\approx 7^{(3)}$              &$\approx 90$          &$1.22 \pm 0.18$                  \\                   
HD\,189733\,b                        &$19.78(0.01)^{(1)}$          &$1.7^{+2.9}_{-0.4}$$^{(4)}$     &$\approx 20$        &$1.14 \pm 0.03$                  \\
$\tau$\,Bootes\,b                   &$15.62(0.05)^{(2)}$             &$\approx 3.2^{(3)}$          &$\approx 100$         &$1.13 \pm 0.17$           \\
$\upsilon$\,Andromedae\,b  &$13.49(0.03)^{(2)}$                 &$\approx 12^{(3)}$        &$\approx 80$           &$1.25 \pm 0.19$                       \\
 \hline
\end{tabular}
\begin{list}{}{}
\item[References:]
$^{(1)}$\citealp{gaia_dr2}; 
$^{(2)}$\citealp{hipparcos}; 
$^{(3)}$\citealp{shkolnik_etal08}; 
$^{(4)}$\citealp{brogi_etal16}; 
$^{(5)}$\citealp{cauley_etal19}.
\end{list}

\end{table}

The capability to detect
low-frequency radio emission from the interaction of an exo-planetary magnetosphere with its 
parent star has been already discussed \citep{zarka07}.
Similarly to the low-frequency radio emission, also the radio
emission at the GHz frequency range, covered by the forthcoming Square Kilometre Array (SKA), 
has been already taken into account \citep{zarka_etal15}.
Exo-planetary auroral radio emission is a bright radio emission
constrained to the low frequency spectral domain, however
at the GHz radio frequencies the SPMI may provide
detectable features too.
Similarly to the auroral radio emission from Jupiter triggered by its moon Io,
an exo-planet moving within the stellar magnetic field
could be able to stimulate auroral radio emission on the magnetic flux tubes connected to the parent star.
In this case the emission frequency of the auroral radio emission depends
upon the stellar magnetic field.
Magnetic field strengths at the kG level are common in case of late type stars \citep{reiners_etal07},
then the auroral radio emission will be tuned at GHz radio frequencies.
Such stellar auroral radio emission triggered by SPMI has been recently discovered
from the star Proxima Cen
\citep{perez_torres_etal21}.
As demonstrated by \citet{zarka07}, both types of SPMI produce 
auroral emission having radio power satisfying  Bode's law.

In accordance with the ECM elementary emission mechanism \citep{melrose_dulk82},
auroral radio emission is a highly directive phenomenon, 
making it difficult to detect despite its strong emission level.
Similarly to the incoherent non-thermal radio emission from Jupiter,
magnetized hot Jupiters might also be direct sources
of such isotropic emission tuned at GHz radio frequencies.
The previous Secs.~\ref{jupiter} and \ref{ucd} support the prediction provided by the scaling relationship
obtained by analyzing the behavior at the radio regime of the early-type magnetic stars and
extrapolated to lower mass objects (UCDs and the planet Jupiter),
for which the detection of their incoherent non-thermal radio emissions have been already reported.
These results stimulated us to adopt the scaling relationship provided in this paper (Eq.~\ref{eq_rel_scal})
as a predictive tool for the incoherent radio emission from astrophysical objects
surrounded by large scale magnetospheres and
having known parameters.
Among these, the hot Jupiter class is suitable.
In fact, some of them are magnetized sub-stellar objects, whose
magnetic field strength, radius, and rotation period
have been constrained.

The favorable condition to produce incoherent non-thermal radio emission
from a large-scale rotating magnetosphere is the corresponding strength of the magnetic field.
In the cases of some hot Jupiters, that are particularly close to the Earth,
the presence of magnetic field strengths higher than Jupiter's
have been reported \citep{cauley_etal19}.
The selected objects are listed in Table~\ref{tab_hot_jup}
with their corresponding parameters.
The expected incoherent radio emission levels
of the four magnetized hot Jupiters,
derived using Eq.~\ref{eq_rel_scal} with the corresponding parameter uncertainties, 
are reported in 
Fig.~\ref{fig_hj} as a function of the effective electromotive force
generated by the rotation of these exo-planets.
The expected fluxes are well below the $\mu$Jy level.

\begin{figure}
\resizebox{\hsize}{!}{\includegraphics{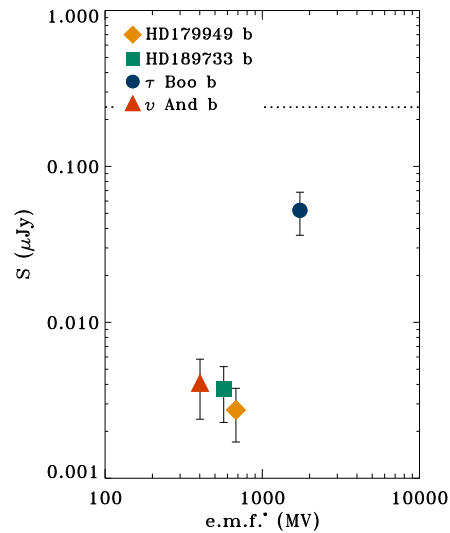}}
\caption{{Expected fluxes at Earth of the four hot Jupiters analyzed.
Fluxes calculated by extending the scaling relationship of the incoherent non-thermal 
radio emission of the early-type magnetic stars. 
The error-bars follow by the parameters uncertainty of the adopted scaling relationship (Eq.~\ref{eq_rel_scal}).
The dotted line represent the SKA detection limit.}}
\label{fig_hj}
\end{figure}

The most sensitive ground based radio interferometer operating at the GHz domain
will be most likely the forthcoming SKA.
In fact the detection limit of the SKA in its full operational state
will be $\approx 0.24$ $\mu$Jy, 
for observations lasting $\approx 10$ minutes of integration time \citep{umana_etal15}.
As seen in Fig.~\ref{fig_hj}, the expected fluxes
of the four hot Jupiters are below the SKA detection limit.
Probably, the hot Jupiters analyzed here do not rotate fast enough,
or are not sufficiently magnetized to provide detectable incoherent radio emission.
In any case, we suspect that in general it is unlikely to expect 
to detect the incoherent radio emission components of hot Jupiters.
In fact, the star's proximity, that is a
suitable condition for triggering the low frequency auroral radio emission,
makes it difficult to spatially resolve its incoherent non-thermal radio emission from the stellar radio emission.

\begin{figure*}
\resizebox{\hsize}{!}{\includegraphics{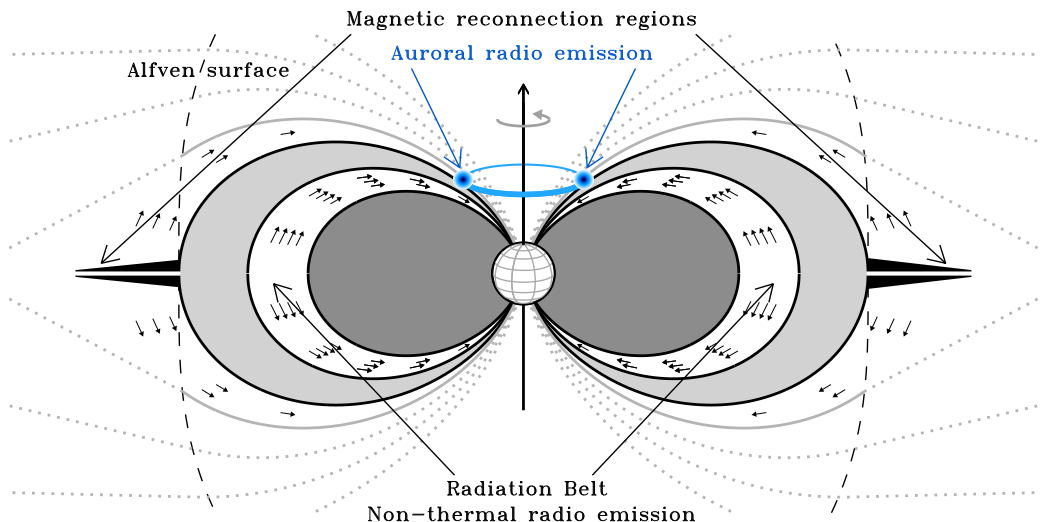}}
\caption{
Schematic representation of the radio emission sites within
a dipole-dominated magnetosphere of a typical early-type magnetic
star.  For this picture the field and rotation axes are
assumed the same.  The location of the radiation belt is highlighted by the
white magnetic shell, with small arrows indicating the relativistic
electrons.  This region is the dominant source of the non-thermal
stellar radio emission, and is located inside the inner magnetosphere.
The cartoon is
purely an idealization, the radiation belt and the Alv\'{e}n surface
are arbitrarily located.  The Alfv\'{e}n surface is far from
the star (dashed line).  Outside this surface the magnetic field
lines are open (dotted lines).  Magnetic reconnection events
take place within the equatorial magneto-disk (dark region), with
consequent production of fast-electrons, that, like the case of Jupiter,
trigger the coherent and strongly beamed auroral radio emission
(blue spots on the blue auroral ring).  
}
\label{scenariofin}
\end{figure*}

The most favorable condition for non-thermal radio emission detection from giant exo-planets is 
the planetary orbital position far from their parent stars, 
making it possible to spatially resolve the radio emissions from both star and exo-planet.
Further, the large distance strongly reduces the star-planet interaction effects,
these ice giant exoplanets might be unable to produce detectable low frequency auroral radio emission,
making the detection of the incoherent non-thermal radio emission 
the only way to search at the radio regime for exoplanets orbiting far from their parent stars.
Non-thermal radio emission may be detectable by SKA in the case of ice giant exo-planets
that, possibly, rotate faster than the tidally locked hot Jupiters and than
produce higher strength magnetic fields,
as expected from the rotation dependent dynamo efficiency.
If the fluxes are above the detection threshold,
the SKA angular resolution in its full operational state, 
$\approx 0.02$ arcsec at $\nu\approx 1$ GHz \citep{umana_etal15},
or lower (depending by the observing frequency),
could be able to easily resolve from their parent stars exo-planets
orbiting at the same distance of Jupiter from the Sun and within 100 pc.
}

\section{Discussion}
\label{discussion}

The observed radio emission of early-type magnetic stars is
compatible with gyro-synchrotron emission from a dipole-like magnetic
shell co-rotating with the star.  The radio spectral {calculations}
allowed us to constrain the equatorial size ($L$ parameter) of the
radiating magnetic shells (Sec.~\ref{sec_subsample}).  Following
the standard scenario for the radio emission from early B magnetic
stars, the shell size was assumed to coincide with the equatorial
Alfv\'{e}n radius, which is the distance where the radiatively driven
stellar wind opens the magnetic field lines.

The inferred wind mass-loss rates required to reproduce the radio spectra 
of the hottest stars are roughly close to (but always higher than) the 
theoretical expectations at the known effective stellar temperatures (Sec.~\ref{sec_wind_comparison}).
For the cooler stars, the above rough accordance totally fails (see Fig.~\ref{wsim_teo}).
The surface chemical composition could play a role in estimation 
of the wind mass-loss rates \citep{krticka14}, or other mass-loss recipes \citep{vink_etal01}
could be employed that predict higher values
(see \citealp{shultz_etal19_490} for details). Ultimately, the 
expected mass loss is still lower than the value required for the 
radio modeling. Independent evidence for a mass-loss inconsistency
at these spectral types comes from HST observations of the wind
sensitive UV lines of the early A-type star HD\,124224
\citep{krticka_etal19}.  Those data allow for only an upper limit
to the wind mass loss, a limit that is discrepant with that required
to explain the observed radio spectrum.

Reconciling the observed common behavior of the radio emission from
the sample of early-type magnetic stars analyzed in
this work, with the discordance between inferred and predicted
mass-loss rates, indicates that we must abandon a physical relation
between the Alfv\'{e}n surface size at the equator ($R_{\mathrm A}$) to
the size of the magnetic shell ($L$) radiating at the radio regime.
Instead, the origin of the non-thermal radio emission must relate
to a physical structure that is less dependent on stellar wind
properties.  We suggest that the non-thermal radio emission originates from radiation belts
located inside the inner-magnetosphere of the early-type magnetic stars (see
Fig.~\ref{scenariofin}).

The stellar wind still plays an important role in
this new picture. The wind is likely the primary source
of ionized matter that fills the stellar magnetosphere.  From
a qualitative point of view, hot stars have favorable
conditions for the onset of all the plasma processes that trigger
the stellar radio emission, simply because their powerful winds 
quickly deposit a large quantity of ionized matter into their
magnetospheres.  Cooler stars are instead characterized by weak
winds and consequently the plasma accumulation is slow.

The radiation belt located inside the inner magnetosphere may be
at a significant distance from the current sheet regions, that form outside
the equatorial Alfv\'{e}n surface, and the site of magnetic reconnection
events, which are the likely sources of electron acceleration.  Now
the question arises: what is the origin of the non-thermal electrons
trapped within the radiation belt responsible for the observed radio
emission?   
One explanation could be
a relation between the origin of the relativistic electrons with
the centrifugal breakout produced by the high-density magnetospheric
plasma locked within the inner magnetosphere.  The breakout mechanism
has already been shown to explain plasma transport within
the centrifugal magnetospheres of the B-type stars
\citep{shultz_etal20,owocki_etal20}.  This is also the case of
Jupiter's magnetosphere, where energetic electrons are produced in
breakdown regions occurring between 15 and 30 $R_{\mathrm J}$
\citep{krupp_etal04}.  Jupiter's radiation belt, located at $L< 5$
$R_{\mathrm J}$, has a closed magnetic topology and is the planet's
source of the non-thermal radio emission. 
For the radio emission from Jupiter's belt zone,
the proposed source of non-thermal electrons 
is the radial diffusion from the farther Jovian magnetospheric regions, where
these energetic electrons are likely produced \citep{bolton_etal04}.

An alternative mechanism for the production of energetic electrons
within closed magnetic field regions is the resonant interaction
of the local plasma with low-frequency electromagnetic waves, known
as whistler modes.  It is this mechanism that seems dominant in the
case of Earth's radiation belt \citep{boyd_etal18}, and also observed
to be active within the radiation belts of the giant planets Jupiter
and Saturn \citep{woodfield_etal14,woodfield_etal19}.  To operate
efficiently, the wave-particle resonant interactions require low
densities, namely $\nu_{\mathrm p} \leq 2.5 \nu_{\mathrm B}$
\citep{horne_etal03}, where $\nu_{\mathrm B}$ is the gyro-frequency
($\nu_{\mathrm B} =2.8 \times B/{\mathrm G}$ MHz) and $\nu_{\mathrm
p}$ is the plasma frequency ($\nu_{\mathrm p} = 9 \times 10^{-3}
\sqrt N_{\mathrm e}$ MHz).  For the early-type stars of this paper,
the typical strength of the magnetic field at distance $L$ is $B_L
\approx 2$ G (Sec.~\ref{sec_spectra_sim}), corresponding to a
gyro-frequency $\nu_{\mathrm B} \approx 6$ MHz.  It follows that
the plasma density of the radiation belt compatible with local
acceleration by wave-particle resonant interaction has to be lower
than $2 \times 10^6$ cm$^{-3}$.

The acceleration through this resonant interaction achieves its
maximum efficiency for whistler modes at an oscillation frequency
of $\nu \approx 0.35 \nu_{\mathrm {B}}$ \citep*{horne_etal03}.  The
corresponding wave frequency of the resonant whistler mode will be
$\nu \approx 2$ MHz, which is lower than the plasma frequency of the Earth's 
ionosphere ($\lessapprox 10$ MHz). This puts strong technical limitations 
on observations below this low-frequency cutoff with ground-based
radio telescopes \citep{de_gasperin_etal18}.  
But such low frequency radiation probably cannot be detected in
any case, even if we observe outside the Earth's ionosphere.

To reproduce the observed radio emission from the early-type magnetic
stars, one must account for the presence of trapped thermal plasma
located within the inner magnetosphere, with a density of order
$10^9$ cm$^{-3}$ close to the stellar surface.  Following the MCWS
model, the density of trapped plasma decreases inversely with distance
from the star.  The plasma surrounding the radiation belt is expected 
to have a density of $\approx 10^8$ cm$^{-3}$. At this value,
low-frequency electromagnetic waves cannot escape from the
stellar magnetosphere. Consequently, wave-plasma interactions
as a proposed electron acceleration mechanism cannot be directly
verified.  Furthermore, some process is needed for forming low-density
cavities within the inner magnetosphere to allow the wave-plasma
interaction to accelerate local electrons efficiently.

In the case of the Earth a correlation of whistlers
mode with the electric discharges related to lightning
\citep{green_etal05}  was observed.  The empirical correlation between the average
radio luminosity of incoherent non-thermal radio emission and
the magnetic flux rate (see Fig.~\ref{lum_mv_mcp}) suggests that
some kind of electrodynamic mechanism may act within the
magnetospheres of the early-type magnetic stars.  The electrodynamical
processes acting within the rotating magnetosphere might be the
engine to originate large scale currents and electric discharges.
Like the case of the Earth, this might be a likely condition to
originate low-frequency electromagnetic waves that, similarly to
the magnetized planets of the solar system, might produce local
plasma acceleration by resonant wave-particle interaction.

Following our new scenario of a radiation belt located within the
inner-magnetosphere, there is a possible critical issue for the
low-frequency coherent auroral radio emission stimulated by the ECM
emission mechanism.  Is this coherent emission capable of escaping
from the inner-magnetosphere to be detectable?  In fact, such
stimulated radio emission originates from the deep magnetospheric
regions close to the star.  Hence, these low-frequency electromagnetic
waves might be strongly suppressed by the gyro-resonant magnetic
absorption of the thermal plasma surrounding the radiation
belt, if the auroral radio emission arises from the same magnetic
shell where the incoherent non-thermal radio emission originates.

The broad studies of Jupiter's radio emission suggest a possible
resolution to the issue raised above.  Jupiter shows clear evidence
of non-thermal radio emission from a radiation belt
located close to the planet ($L < 5$ R$_\mathrm{J}$), 
and also various components of coherent radio emission.  
An important source of coherent emission is associated with the moon Io, but
there is also a non-Io contribution arising from the magnetic field
lines anchored at the planet's surface at high latitudes, and
coinciding with the main auroral oval.  It was recognized that the
footprints of these field lines are related to the magnetic field
lines crossing the magnetic equatorial plane at a large distance ($L
\gtrapprox 7$--15 R$_\mathrm{J}$) and close to the breakdown regions
where magnetic reconnection events have been identified.  
These
reconnection events provide the injection of fast electrons that
trigger {on Jupiter} the non-Io DAM {(i.e. the component of the Jovian decametric radio emission (DAM)
not related to the magnetic flux tube connected with the moon Io)}
and the bKOM {(i.e. the Jovian broadband kilometric radiation) 
\citep{zarka98,zarka04}.
The Jupiter-like scenario was recently supported to be scalable also in the case of magnetic early type stars
by the frequency dependent behavior 
of the wide band coherent emission 
from HD\,124224 (CU\,Vir), the prototype of the class of early-type magnetic stars showing auroral radio emission,
where several auroral radio emission components
seems to be related to different magnetic field lines of the stellar magnetosphere
\citep{das_chandra21}.}

Similar to Jupiter, the auroral radio emission of early-type
magnetic stars might be triggered by the non-thermal electrons
accelerated within the magnetospheric regions where the magnetic
field strength is too weak to confine the magnetospheric wind plasma, or
produced by continuous centrifugal breakouts events.
The result is the formation of an equatorial magneto-disk site of
magnetic reconnection.  A cartoon that summarizes this scenario
is given in Fig.~\ref{scenariofin}.
{Even if this scenario is an over-simplified schematization of a true rotating stellar magnetosphere, 
this picture highlights that the injection of non-thermal electrons is not related to a unique acceleration site, 
where, possibly, different acceleration mechanisms are working. 
Beyond the magnetospheric regions where the wind plasma breaks the field lines, 
producing magnetic reconnection and consequent acceleration of electrons,
energetic electrons can be also locally accelerated within the most internal regions 
where the magnetic field lines are closed, as seen in the case of Earth.
}

\section{Conclusion and outlook}
\label{sec_conclusion}

In this paper, we analyzed the incoherent gyro-synchrotron radio emission of 
a sample of early type magnetic stars, ranging from early-B to early-A type stars,
as a function of their stellar parameters. As a result of our 
study, we report the existence of a strict correlation between the 
non-thermal radio luminosity ($L_{\nu,\mathrm{rad}}$) and the ratio 
of magnetic flux to the rotation period ($\Phi/P_{\mathrm{rot}}$),
given by the Eq.~\ref{eq_rel_scal} and shown in Fig.~\ref{lum_mv_mcp}.
{This relationship appears to be something analogous to the scaling relationship
that describes the coherent auroral radio emission of
the magnetized body of the Solar system, formalized 
by the radio-magnetic Bode's law and which is extrapolated up to the exo-planets \citep{zarka07}.}

{We also reviewed the wind scenario used to account for 
the non-thermal radio emission of these stars. 
{The stellar wind has still a key role
to supply ionized material within the magnetospheres of the early-type magnetic stars.
But, unlike what was previously assumed,
we found that the equatorial size of the magnetospheric region radiating incoherent gyro-synchrotron radio emission
is not large enough to intercept the regions in which the wind pressure equals the magnetic tension.
To explain the features of the non-thermal radio emission from such hot magnetized stars we proposed}
the existence of a Jupiter-like 
radiation belt  
within the magnetospheres of the 
early-type magnetic stars. 
{The existence of a radiation belt around the magnetized planets of the Solar system
is a well-known phenomenon,
already taken into account in stellar cases too.
To explain the origin of the incoherent non-thermal radio emission 
from UCDs, a plasma radiation belt has been 
proposed to exist within their magnetospheres
\citep{hallinan06}.
This idea applied in the context of the early type magnetic stars 
leads to convenient results.
In fact, in}
this new picture, the magnetic shell 
where the non-thermal radio emission originates is formally decoupled 
from the wind mass loss rate {strength.
Following this new wind paradigm, a more complex picture seems to come out 
with regard to the origin of the non-thermal electrons responsible for the incoherent gyro-synchrotron emission.
In particular, the current sheets produced by the wind magnetic breakout
can not provide unique (and primary) acceleration sites.}

Importantly, the scaling relationship
between $L_{\nu,\mathrm{rad}}$ and $\Phi/P_{\mathrm{rot}}$ for the 
early-type magnetic stars, also successfully predicts 
the non-thermal radio emission from some (six) UCDs and the planet Jupiter.
The parameter $\Phi/P_{\mathrm{rot}}$ is related to the effective 
electromotive force generated by the rotation of a large-scale and 
well-ordered magnetic field. The rotation stimulates electrodynamic 
processes occurring within the stellar magnetosphere, that appear 
key to establishing the radio emission level of 
the dipole like magnetospheres surrounding objects 
at both the top and at the bottom of 
the main sequence (see Fig.~\ref{lum_mv}). 
It is then possible that the mechanism for 
acceleration of non-thermal electrons may be related to large-scale 
stationary current systems.

The various {plasma configurations
that we have suggested as possibly existing}
within the magnetospheres of early-type stars certainly
demands further investigation. And while we have identified a
scaling relation that span magnetospheres from early-type massive
stars to UCDs and planets, this needs further study. 
{The early-type magnetic stars have a favorable condition 
to fill their magnetospheres with ionized material.
In fact, the stellar wind, even if in some cases the mass-loss rate is at a very weak level,
has a key role as a plasma source.}
The UCDs
and Jupiter have well-ordered dipolar fields, but are
cold objects lacking winds of any significance, yet they produce
non-thermal radio emission.  
However, if some other plasma source fills their magnetospheres
(i.e. exo-planetary volcanic activity analogous to that of the moon Io), 
suitable conditions for the radio emission generation will be realized.
The radio luminosity of 
incoherent emission from UCDs is low, and for only a few 
such emission
has been detected, while
simultaneously having reliable magnetic field strengths.  
Nonetheless,
the few cold objects that we have analyzed display radio behavior
in remarkable accord with the empirical scaling relationship deduced 
from the early-type stars.  
This is a clear
hint of a common behavior among both stars and substellar objects
surrounded by large-scale and stable magnetospheres, suggesting
that an underlying and unifying physical mechanism is at work.  In
particular, it is plausible to hypothesize the existence of a
common acceleration mechanism of electrons to relativistic
energies operating within the magnetospheres among these extremely
different classes of objects.

Future work must enlarge the sample of early-type magnetic stars 
to refine the empirical relationship  
for the non-thermal radio emission from ordered magnetospheres.
In particular, a systematic search 
for radio emission at the top of the main sequence should be 
useful for the indirect estimation of stellar parameters, in particular 
the stellar magnetic field strength. In fact, on the basis of the 
empirical dependence between radio luminosity and stellar parameters,
the detection of non-thermal radio emission might be used to constrain
the stellar magnetic field strength.

The next generation of ground based radio interferometer (e.g. SKA) 
will quickly be able to reach {a detection limit at}
the $\mu$Jy
level, such that
brighter early-type magnetic stars ($L_{\nu,{\mathrm{rad}}} \approx 
10^{18}$ ergs\,s$^{-1}$\,Hz$^{-1}$) located close to the Galactic 
center ($\approx 9$ kpc distance) will be easily detected 
(expected flux levels $\approx 10$ $\mu$Jy). Ap/Bp stars 
are uniformly distributed in space \citep{renson_manfroid09}, and 
the number recognized as magnetic stars is 
rapidly increasing \citep{fossati_etal15,wade_etal16}.
At the present, spectropolarimetric measurements are able to 
perform reliable measurements of stellar magnetic fields only for stars 
located within 1--2 kpc from the Sun \citep{shultz_etal19_485,bagnulo_etal20}. We 
emphasize that radio emission could be used as another diagnostic 
tool for stellar magnetism, and leveraged to investigate the incidence 
of magnetism at the top of the main sequence. The origin of 
magnetic fields of early-type stars, and their persistence 
over their MS life, and beyond \citep{Neiner2017, Martin2018}, is a crucial matter. In fact, 
the magnetic field of the very massive
stellar progenitor might be significant 
to the core-collapse supernovae explosion and the potential birth 
of magnetars \citep{bucciantini_etal09,kasen_bildsten10}.

\section*{Data availability}
The data underlying this article are available in the article and in its appendices.

\section*{Acknowledgments}
We sincerely thank the anonymous referee for his/her very useful and constructive criticisms and suggestions.
This work has extensively used the NASA's Astrophysics Data System, and the 
SIMBAD database, operated at CDS, Strasbourg, France. 
LMO acknowledges support from
the DLR under grant FKZ\,50\,OR\,1809 and partial support by the Russian
Government Program of Competitive Growth of Kazan Federal University.
FL was supported by ``Programma ricerca di ateneo UNICT 2020-22 linea 2''.
RI acknowledges funding support for this research from a grant
by the National Science Foundation, AST-2009412.
MG acknowledges financial contribution from the agreement ASI-INAF n.2018-16-HH.0.
MES acknowledges financial support from the Annie Jump Cannon Fellowship, supported by the University of Delaware and endowed by the Mount Cuba Astronomical Observatory.
A special thank to the project MOSAICo 
(Metodologie Open Source per l'Automazione Industriale e delle procedure di CalcOlo in astrofisica) funded by Italian MISE (Ministero Sviluppo Economico).




\appendix

\section{Radio measurements of the individual early type stars}
\label{radio_appendix}

\subsection{HD\,12447}
\label{radio_12447}
This magnetic Ap star was detected at 1.4 GHz by the FIRST (Faint Images of the Radio Sky at Twenty centimeters) survey performed by the VLA \citep{becker_etal95},
the flux  is $S=0.7  \pm 0.1$ mJy \citep{helfand_etal15}. 
The stellar radio emission was also detected at 3.6 cm (corresponding to $\approx 8.4$ GHz) with a flux level of
$S=0.37\pm0.09$ mJy \citep{drake_etal06}.
Further, HD\,12447 holds in the tile T11t04 observed at 3 GHz in two epoch of the VLA Sky Survey (VLASS1.1 and 2.1)
and was also detected at both epochs with similar flux level, on average $0.69\pm0.09$ mJy.
The radio luminosity of  HD\,12447 reported in Table~\ref{tab_stars} 
has been obtained using 
the average fluxes at these available radio frequencies.
Mean frequency
$<\nu>\approx 3.3$ GHz.

\subsection{HD\,19832}
\label{radio_19832}
HD\,19832 was detected by the VLA at $8.4$ GHz, $S=0.45\pm0.12$ mJy \citep*{drake_etal06},
and at 3 GHz by the second half of the first epoch (VLASS1.2, tile T17t05) of the VLASS ($S=0.5\pm0.1$ mJy).
Also in this case, the radio luminosity reported in Table~\ref{tab_stars} 
has been retrieved by
the average flux of the two radio measurements, at the corresponding mean frequency
$<\nu>\approx 5$ GHz.

\subsection{HD\,27309}
\label{radio_27309}
Measurements of the radio emission of HD\,27309
are available at three distinct frequencies,
respectively at $\nu=3$ GHz (VLASS1.2, tile T16t06), $S=0.4\pm0.1$ mJy, and
at 5, $S=0.38\pm0.02$ mJy, and 8.4 GHz, $S=0.33\pm0.02$ mJy \citep*{drake_etal06}.
To estimate the radio luminosity of HD\,27309 (Table~\ref{tab_stars}), 
the average flux has been used.
Mean frequency
$<\nu>\approx 5$ GHz.

\subsection{HD\,34452}
\label{radio_34452}
The detection of the radio emission of IQ\,Aur (HD\,34452) at $\nu=5$ GHz ($S=0.48\pm0.09$ mJy) 
was firstly reported by \citet{drake_etal87}
and then confirmed by \citet*{linsky_etal92} ($S=0.32\pm0.05$ mJy).
The average flux has been used to estimate the radio luminosity of HD\,34452. 

\subsection{HD\,35298}
\label{radio_35298}
It was recently discovered by \citet{das_etal19} that HD\,35298 is a source of low frequency ($\nu \leq 1.4$ GHz) 
high strength (flux level approaching up to $\approx10$ mJy) coherent auroral radio emission.
Such maser emission is constrained within a well defined range of phases 
and is superimposed above the basal emission produced by
the incoherent gyro synchrotron mechanism.
The incoherent component of the HD\,35298 emission was firstly measured at 5 GHz by \citet*{linsky_etal92}
($S=0.28\pm0.06$ mJy). 
Measurements at other frequencies are also available in the literature
($S=0.29\pm0.07$ mJy at $\nu=8.4$ GHz; \citealp*{drake_etal06}).
Further, our unpublished high frequencies VLA observations (project code: AL618)
allowed us to measure the fluxes  of HD\,35298 at $\nu=15$ GHz ($S=0.32\pm0.03$ mJy)
and at $\nu=22$ GHz ($S=0.13\pm0.02$ mJy). 
These radio measurements cover a wide frequency range and
have been averaged to estimate the luminosity reported in Table~\ref{tab_stars} 
at the
mean frequency $<\nu>\approx 10.9$ GHz.

\subsection{HD\,35502}
\label{radio_35502}
HD\,35502 holds in the tile T10t09 observed over two epochs of the VLA Sky Survey at 3 GHz (VLASS1.1 and 2.1),
with fluxes $S=2.6\pm0.1$ and $S=3.0\pm0.1$ mJy.
Further, unpublished multi-wavelengths VLA radio observations of HD\,35502 have been retrieved querying the
NRAO data archive (project code: AL388).
This data set has been processed by us to obtain the following radio measurements:
$S=1.7\pm0.1$ mJy at $\nu=5$ GHz;
$S=2.28\pm0.07$ mJy at $\nu=8.4$ GHz;
$S=1.36\pm0.06$ mJy at $\nu=15$ GHz.
Another measurement performed at 8.4 GHz is reported in the literature.
The flux of HD\,35502 at that observing epoch was $S=2.97\pm0.1$ mJy \citep*{drake_etal06}.
These multi-frequencies measurements are used to obtain the average radio luminosity listed in 
Table~\ref{tab_stars}.
Mean frequency
$<\nu>\approx 6.6$ GHz.

\subsection{HD\,36485}
\label{radio_36485}
\begin{table}
\caption[ ]{HD\,36485 radio spectrum.}
\label{tab_36485}
\footnotesize
\begin{tabular}{c c l}
\hline
$\nu$                 &$<S_{\nu}>$                          &                \\
(GHz)                &(mJy)                        &notes                          \\
\hline       
$0.61^{\ast}$  &$0.6\pm0.2$       & \citet{chandra_etal15}               \\  
$1.4$               &$0.8\pm0.2$            & \citet{chandra_etal15}; AL267$^{\dag}$   \\
$3^{\ast}$        &$1.4\pm0.2$             & VLASS1.1 and 2.1 (tile T10t09)              \\
$5$                  &$1.1\pm0.2$              & \citet{leone_etal10}                     \\  
$8.4$               &$1.0\pm0.2$             & AL267$^{\dag}$; AL388                     \\  
$15$                &$0.8\pm0.2$               & AL267$^{\dag}$; AL388                      \\  
\hline
\end{tabular}
\begin{list}{}{}
\item[] $^{\ast}$ Less than three measurements. $^{\dag}$ NRAO image archive.
\end{list}
\end{table}
The radio emission of HD\,36485 was firstly reported by \citet{drake_etal87}.
HD\,36485 was observed many times at $\nu=5$ GHz.
These radio observations cover the whole stellar rotation period. 
The light curve at 5 GHz was performed by \citet{leone_etal10} and
shows a moderate rotational modulation. 
Additional radio observations of HD\,36485, performed by the VLA at other frequencies, exists.
In this paper, we analyzed all the available data set.
The VLA project AL267, beyond the already published 5 GHz radio measurements, 
contains observations performed at $\nu=1.4$, 8.4 and 15 GHz, performed at three different epochs (1992-Oct-11, 13, and 15).
These multi-frequency observations have been processed by the standard VLA pipeline. 
The radio maps are available querying the NRAO Image Retrieval Tool\footnote{\url{https://archive.nrao.edu/archive/archiveimage.html}}
and are used for measuring the stellar radio fluxes.
Further unpublished VLA observations at 8.4 and 15 GHz are available (project AL388).
These archival data have been reduced by us using the standard calibration procedure enabled in {\sc casa}.
Ancillary radio measurements at 0.61 and 1.4 GHz are also collected \citep{chandra_etal15}.
Finally, HD\,36485 was also detected at 3 GHz by the two epochs of the first half VLASS (VLASS1.1 and 2.1).
All the available radio measurements performed at the same frequency
have been averaged, listed in Table~\ref{tab_36485}, and shown in Fig.~\ref{spe10sou}.
In the case of more than one measurement at the same frequency,
the uncertainty has been estimated adding in quadrature the measurements dispersion
and the average of the individual errors.
The data in Table~\ref{tab_36485} has been averaged to estimate the stellar radio luminosity
at the mean frequency $<\nu>\approx 3.4$ GHz.

\subsection{HD\,37017}
\label{radio_37017}
\begin{table}
\caption[ ]{HD\,37017 radio spectrum.}
\label{tab_37017}
\footnotesize
\begin{tabular}{c c l}
\hline
$\nu$                 &$<S_{\nu}>$                          &                \\
(GHz)                &(mJy)                        &notes                          \\
\hline       
$0.61^{\ast}$   &$0.6\pm0.3$           & (a)               \\  
$1.4$               &$1.6\pm0.5$            & (a); (b); (c)   \\
$3^{\ast}$        &$1.7\pm0.7$             & VLASS1.1 and 2.1 (tile T09t09)              \\
$4.5$                  &$2.0\pm0.2$              & {(d)}                  \\  
$5$                  &$2.0\pm0.5$              & (b); (e); (f)                   \\  
$8.4$               &$2.1\pm0.3$             & AL267$^{\dag}$; AL388                      \\  
$15$                &$1.8\pm0.4$              & (b); (c); AL267$^{\dag}$; AL388; AL618                      \\  
$22$                &$2.0\pm0.9$              &  (c); AL234$^{\dag}$; AU53$^{\dag}$; AL618               \\  
$43^{\ast}$     &$1.9\pm0.1$              & AL618                      \\  
\hline
\end{tabular}
\begin{list}{}{}
\item[] $^{\ast}$ Less than three measurements. $^{\dag}$ NRAO image archive.
(a) \citealp{chandra_etal15};
(b) \citealp{drake_etal87}; 
(c) \citealp{leone_etal96};
{(d)} \citealp{kounkel_etal14}.
(e) \citealp{leone_umana93};
(f) \citealp{trigilio_etal04}.

\end{list}
\end{table}
The radio emission of HD\,37017 was firstly measured by \citet{drake_etal87} and then
the rotation modulation effect on the 5 GHz stellar radio emission was analyzed \citep{leone91,leone_umana93}.
The 5 GHz light curve of HD\,37017 was  {calculated} 
by \citet{trigilio_etal04}
and compared with the available measurements covering the stellar rotation period.
The average value of these measurements is listed in Table~\ref{tab_37017}.
{ A reliable VLA detection ($\nu=4.5$ GHz) has been also reported by \citet{kounkel_etal14}}
The tentative detection of the low-frequency ($\nu=610$ MHz) emission of HD\,37017 was reported by \citet{chandra_etal15},
further, measurements at other radio frequencies are also reported in the literature.
The ancillary data complemented by unpublished measurements (reduced by us or processed by the standard VLA pipeline)
acquired at the same frequency are averaged altogether.
Further, HD\,37017 was detected at 3 GHz by the VLASS 
(in the table the average of the two epochs of the first half sky survey, respectively VLASS1.1 and VLASS2.1, was reported).
The average spectrum reported in Table~\ref{tab_37017} is pictured in Fig.~\ref{spe10sou}.
The mean flux value has been used to estimate the radio luminosity of HD\,37017.
Mean frequency
$<\nu>\approx 5.8$ GHz.

\subsection{HD\,37479}
\label{radio_37479}
The non-thermal radio emission of HD\,37479 was firstly measured by \citet{drake_etal87}.
This star has been then subject of many radio observations.
A detailed study of the multi-frequency radio emission of HD\,37479
was performed by \cite{leto_etal12}, where a great number of radio measurements are reported.
{
The radio survey of the Orion region performed by the VLA at 4.5 and 7.5 GHz \citep{kounkel_etal14}
also detected the radio emission of HD\,37479.}
\citet{chandra_etal15} 
explored the HD\,37479 behavior at the low frequency spectral region ($\nu=610$ MHz).
HD\,37479 was also detected at 3 GHz by the VLASS in two epochs (1.1 and 2.1).
The data available at each radio frequency have been averaged,
this average radio spectrum is reported in Table~\ref{tab_37479} and pictured in Fig.~\ref{spe10sou}.
The reported errors are the sum in quadrature of the dispersions of the data acquired at the same frequency 
and of the average of the uncertainties of the individual measurements.
The mean value of these multi-frequencies radio measurements has been used to estimate the 
radio luminosity at the mean frequency $<\nu>\approx 5.9$ GHz.

\subsection{HD\,40312}
\label{radio_40312}
Two distinct detections with similar fluxes of the stellar radio emission at 5 GHz exist  \citep*{linsky_etal92}.
It was commonly observed the rotational modulation of the radio emission from the hot magnetic stars.
To take into account the expected radio variability,
we assumed the uncertainty of 50\%.
The measured flux ($S\approx0.3$ mJy)
has been used to estimate the radio luminosity of HD\,40312. 

\begin{table}
\caption[ ]{HD\,37479 radio spectrum.}
\label{tab_37479}
\footnotesize
\begin{tabular}{c c l}
\hline
$\nu$                 &$<S_{\nu}>$                          &               \\
(GHz)                &(mJy)                        &notes                          \\
\hline       
$0.61^{\ast}$  &$1.2\pm0.3$         & (a)               \\  
$1.4$               &$2.3\pm0.5$            & (a); (b); (c)   \\
$3^{\ast}$        &$3.0\pm0.2$             & VLASS1.1 and 2.1 (tile T10t09)              \\
$4.5$                  &$3.7\pm0.1$              & {(d)}                 \\  
$5$                  &$3.8\pm0.5$              & (c)                     \\  
$7.5$                  &$3.8\pm0.2$              & {(d)}                 \\  
$8.4$               &$3.8\pm0.6$             & (c)                     \\  
$15$                &$4.0\pm0.9$              & (c)                      \\  
$22$                &$3.7\pm0.9$              & (c)                      \\  
$43$                &$3.1\pm0.9$              & (c)                     \\  
\hline
\end{tabular}
\begin{list}{}{}
\item[] $^{\ast}$ Less than three measurements.
(a) \citealp{chandra_etal15};
(b) \citealp*{linsky_etal92}; 
(c) \citealp{leto_etal12};
{(d)} \citealp{kounkel_etal14}.

\end{list}
\end{table}

\subsection{HD\,79158}
\label{radio_79158}
Only one detection of the radio emission from HD\,79158 exists.
This star was observed at 8.4 GHz and detected with the flux of $S=0.45$ mJy \citep*{drake_etal06}.
To take into account possible rotational modulation,
we assumed a flux uncertainty of 50\%.
This single data has been used to estimate the radio luminosity of HD\,79158. 

\subsection{HD\,112413}
\label{radio_112413}
This is the prototype of the class of variable stars of $\alpha^2$\,CVn type.
HD\,112413 ($\alpha^2$\,CVn) was detected at 1.4 GHz by the FIRST survey
($S=0.77 \pm  0.09$ mJy; \citealp*{becker_etal95}).
This star was also detected at 8.4 GHz ($S=0.29$ mJy; \citealp*{drake_etal06})
and at 3 GHz by the VLASS.
The source was detected by both epochs of the first half sky survey
(VLASS1.1 and 2.1, tile T20t17), the average flux is $S=1.2 \pm  0.2$ mJy.
These multi-frequency detections have been averaged to estimate
the radio luminosity of HD\,112413 (Table~\ref{tab_stars}). 
Mean frequency
$<\nu>\approx 3.3$ GHz.

\subsection{HD\,118022}
\label{radio_118022}
HD\,118022 is the first star, which is not the Sun, in which the effect of the magnetic field was revealed in the stellar spectra \citep{babcock47}.
Its radio emission was detected  by the second epoch of the VLA Sky Survey (VLASS1.2),
tile T11t21, the measured flux is $S = 0.5$ mJy.
This star was already observed by the VLA in the early 1990s at 8.4 GHz (project AD288), 
but the poor quality image prevents us from detecting its radio emission.
Also in this case, it was assumed the flux uncertainty of 50\% to estimate the radio luminosity of HD\,118022.

\subsection{HD\,124224}
\label{radio_124244}
\begin{table}
\caption[ ]{HD\,124224 radio spectrum.}
\label{tab_124224}
\footnotesize
\begin{tabular}{c c l}
\hline
$\nu$                 &$<S_{\nu}>$                          &               \\
(GHz)                &(mJy)                        &notes                          \\
\hline       
$1.4$               &$2.5\pm0.6$            & (a); (b); (c); (d)   \\
$2.5$               &$3.2\pm0.4$            & (c)   \\
$3^{\ast}$        &$2.7\pm0.2$             & VLASS1.2 (tile T11t22)              \\
$5$                  &$3.9\pm0.6$              & (e)                     \\  
$8.4$               &$3.9\pm0.6$             & (e)                     \\  
$15$                &$3.4\pm0.6$              & (e)                      \\  
$22^{\ast}$                &$3\pm1$                   & (a); AL618                      \\  
$87.7^{\ast}$  &$1.3\pm0.3$              & (e)              \\  
\hline
\end{tabular}
\begin{list}{}{}
\item[] $^{\ast}$ Less than three measurements.
(a) \citealp*{leone_etal96};
(b) \citealp{trigilio_etal00};
(c) \citealp{trigilio_etal08}; 
(d) \citealp*{helfand_etal15}; 
(e) \citealp{leto_etal06}; 
(f) \citealp{leone_etal04}.
\end{list}
\end{table}
This is the first star where the coherent auroral radio emission was observed \citep{trigilio_etal00}
and is widely studied at the radio regime \citep{leto_etal06}.
In this paper we collected many already published multi frequency radio measurements.
Further, we analyzed our unpublished high-frequency VLA observations (project code AL618),
data analyzed and imaged using {\sc casa}.
The radio measurements performed at the same radio observing band were averaged altogether,
the related flux uncertainty is the sum in quadrature of the measure dispersion and the mean value of the individual errors.
The average spectrum of HD\,124224 is pictured in Fig.~\ref{spe10sou}.
The mean value of the data listed in Table~\ref{tab_124224}  has been used to estimate the radio luminosity of HD\,124224.
Mean frequency $<\nu>\approx 7.7$ GHz.

\subsection{HD\,133652}
\label{radio_133652}
One epoch VLA observation at 8.4 GHz of HD\,133652 exists.
A tentative detection of the stellar radio emission ($S=0.24$ mJy) is reported by \citet*{drake_etal06}.
Also in this case we adopted the flux uncertainty of  50\% 
to estimate the radio luminosity of HD\,133652.

\subsection{HD\,133880}
\label{radio_133880}
HD\,133880 was firstly observed by the ATCA interferometer covering the whole stellar rotation period \citep{lim_etal96}.
The above data set allowed to perform the light curves at 5 and 8.4 GHz \citep{bailey_etal12}.
At lower radio frequencies HD\,133880 evidenced enhanced emission constrained in a narrow range of phases \citep{chandra_etal15},
later explained as coherent auroral radio emission \citep*{das_etal18}.
Further, the wide band analysis of the auroral emission from HD\,133880 allowed to evidence the
longitudinal dependence of the plasma accumulation within the stellar magnetosphere \citep{das_etal20b}.
Despite its low sky latitude ($\delta=-40.5839^{\circ}$) this star is visible also by the VLA.
The archival data set AD348 contains observations performed at three different epochs (1995-Feb-15, 16, and 17),
the images are available querying the NRAO image archive.
Further, HD\,133880 was detected also at 43 GHz (VLA data set AL388 reduced and imaged in this paper).
The low frequencies measurements ($\nu=0.61$ and 1.4 GHz; \citealp{chandra_etal15}) 
acquired at phases where the auroral radio emission of HD\,133880 is not observable have been 
used to estimate the incoherent emission component. 
Also in the case of measurements at the higher frequencies, the data at the same frequency have been averaged.
The mean radio spectrum is reported in Table~\ref{tab_133880} and pictured in Fig.~\ref{spe10sou}. 
The average flux of HD\,133880 performed using all the available frequencies 
has been used to estimate its radio luminosity at the
mean frequency $<\nu>\approx 5.3$ GHz.

\begin{table}
\caption[ ]{HD\,133880 radio spectrum.}
\label{tab_133880}
\footnotesize
\begin{tabular}{c c l}
\hline
$\nu$                 &$<S_{\nu}>$                          &               \\
(GHz)                &(mJy)                        &notes                          \\
\hline       
0.61                &$2.2\pm0.7$         & \citet{chandra_etal15}              \\  
$1.4$               &$3\pm1$            &  \citet{chandra_etal15}   \\
$5$                  &$3.5\pm0.8$              & \citet{bailey_etal12}                     \\  
$8.4$               &$3.6\pm0.6$             & \citet{bailey_etal12}                     \\  
$15$                &$3.2\pm0.5$              & AD348$^{\dag}$       \\  
$43^{\ast}$     &$1.0\pm0.2$              & AL388              \\  
\hline
\end{tabular}
\begin{list}{}{}
\item[] $^{\ast}$ Less than three measurements. $^{\dag}$ NRAO image archive.
\end{list}
\end{table}

\subsection{HD\,142184}
\label{radio_142184}
\begin{table}
\caption[ ]{HD\,142184 radio spectrum.}
\label{tab_142184}
\footnotesize
\begin{tabular}{c c l}
\hline
$\nu$                 &$<S_{\nu}>$                          &               \\
(GHz)                &(mJy)                        &notes                          \\
\hline       
$0.8875^{\ast}$  &$6.2\pm0.2$           &  \citet{pritchard_etal21}   \\
$1.4^{\ast}$    &$12.6\pm0.6$            &  \citet{condon_etal98}   \\
$3^{\ast}$        &$30\pm2$             & average of VLASS1.1 and 2.1 (tile T05t24)              \\
$6$               &$73\pm15$             & \citet{murphy_etal10} ; \citet{leto_etal18}                      \\  
$8^{\ast}$         &$95\pm5$              & \citet{murphy_etal10}                     \\  
$10$               &$100\pm20$             & \citet{leto_etal18}                      \\  
$15$                &$110\pm20$              & \citet{leto_etal18}                       \\  
$20^{\ast}$         &$104\pm6$              & \citet{murphy_etal10}                     \\  
$22$                &$120\pm30$                   & \citet{leto_etal18}                      \\  
$33$                &$130\pm30$                   & \citet{leto_etal18}                      \\  
$44$                    &$120\pm40$              & \citet{petrov_etal12,leto_etal18}              \\  
$102^{\ast}$       &$115\pm20$              & \citet{leto_etal18}              \\  
$292^{\ast}$      &$32.5\pm3$              & \citet{leto_etal18}              \\  
\hline
\end{tabular}
\begin{list}{}{}
\item[] $^{\ast}$ Less than three measurements.
\end{list}
\end{table}
HD\,142184 is a bright radio source detected up to the millimeter range,
in fact, it was detected by the ALMA telescope \citep{leto_etal18}.
The measurement at the lowest available frequency ($\nu=887.5$ MHz) has been performed by the Australian SKA Pathfinder (ASKAP) interferometer  \citep{pritchard_etal21}.
Almost all the available radio measurements of this magnetic star have been here collected. 
The fluxes acquired at similar observing frequencies have been averaged, these are listed in Table~\ref{tab_142184}.
At the lower frequencies
the spectral dependence of the polarization fraction of the incoherent emission from HD\,142184  
evidenced a decaying trend, the polarization fraction decreases from a level of $\approx 10$ \% at 44 GHz down to a few \% at 6 GHz  \citep{leto_etal18},
this is in accordance with the canonical incoherent gyro-synchrotron emission mechanism.
The low frequency ASKAP data ($\nu=887.5$ MHz) presents a circular polarization fraction of $\approx 22$\%.
This high level of circular polarization fraction observed at $887.5$ MHz is incompatible with the canonical spectral dependence of the 
circularly polarized emission from the incoherent gyro-synchrotron mechanism,
this is a highly plausible hint that also the coherent auroral radio emission contributes to the low frequency radio emission from HD\,142184.
Then, the flux at $\nu=0.8875$ GHz listed in Table~\ref{tab_142184} is the total intensity given by \cite{pritchard_etal21}
reduced of the 22\%, 
assumed as the auroral coherent emission fraction contaminating the incoherent non-thermal emission at $\nu=0.8875$ GHz.
The mean spectrum is pictured in Fig.~\ref{spe10sou}. 
The average radio luminosity of HD\,142184, at the mean frequency $<\nu>\approx 13.5$ GHz, is reported in Table~\ref{tab_stars}. 

\subsection{HD\,142301}
\label{radio_142301}
\begin{table}
\caption[ ]{HD\,142301 radio spectrum.}
\label{tab_142301}
\footnotesize
\begin{tabular}{c c l}
\hline
$\nu$                 &$<S_{\nu}>$                          &               \\
(GHz)                &(mJy)                        &notes                          \\
\hline       
$1.5$               &$2.8\pm0.4$            &  \citet{leto_etal19}   \\
$3^{\ast}$        &$2.6\pm0.2$             & VLASS1.2 (tile T04t24)              \\
$5$                  &$4.0\pm0.7$              & \citet{leto_etal19}                     \\  
$8.4$               &$3.8\pm0.9$             &  \citet*{linsky_etal92}; AL388                     \\  
$15^{\ast}$     &$3.4\pm0.5$              & \citet*{leone_etal96}; AL388       \\  
$43^{\ast}$     &$2.9\pm0.2$              & AL388              \\  
\hline
\end{tabular}
\begin{list}{}{}
\item[] $^{\ast}$ Less than three measurements. 
\end{list}
\end{table}

Multifrequency measurements of the incoherent non-thermal radio emission of
HD\,142301 are reported by \citet*{linsky_etal92} and by \citet*{leone_etal96}.
HD\,142301 is also a source of strongly polarized coherent auroral radio emission \citep{leto_etal19}.
Such kind of maser emission is constrained to a narrow range of the stellar rotational phases.
To estimate the incoherent contribution alone, we average the 1.5 and the 5 GHz measurements acquired at phases 
far from the coherent pulses. 
Further, HD\,142301 was detected by the VLASS at 3 GHz and
unpublished VLA observations (project AL388) also revealed the high frequency contribution of the incoherent stellar radio emission
(data set reduced and analyzed in this paper).
The radio measurements acquired at the same frequency range have been averaged,
the average spectrum is reported in Table~\ref{tab_142301} and shown in Fig.~\ref{spe10sou}.
The mean flux is used to estimate the stellar radio luminosity at the
mean frequency $<\nu>\approx 7$ GHz.

\subsection{HD\,142990}
\label{radio_142990}
The confirmed evidence of coherent auroral radio emission from HD\,142990 was reported by
\citet{das_etal19a}, which observed this star at low frequency ($\nu\approx 600$ MHz and $\nu \approx 1.4$ GHz) using the GMRT.
At these low frequencies, the auroral emission from HD\,142990 was observed in form of strongly polarized large pulses,
that strongly contaminate the basal incoherent emission component.
This condition make unreliable the estimation of the low-frequency incoherent emission component,
that was roughly estimated at the mJy level.
The incoherent emission component of HD\,142990 was instead clearly measured at higher frequency bands.
The average flux of HD\,142990 measured at 5 GHz by \citet*{linsky_etal92},
by \citet*{leone_etal94}, and by us, analyzing archival VLA data (project AL388), is
$S=1.9 \pm 0.3$ mJy.
A similar flux level was observed at 8.4 GHz ($S=1.8 \pm 0.3$ mJy; \citealp*{linsky_etal92} and VLA project AL388).
The mean flux between the 5 and 8.4 GHz measurements has been used to estimate the stellar radio luminosity. 
Mean frequency
$<\nu>\approx 6.5$ GHz.

\subsection{HD\,144334}
\label{radio_144334}
The radio emission of HD\,144334 at $\nu=5$ GHz was detected in two different epochs by \citet*{linsky_etal92}.
The average flux is $S=0.3 \pm 0.15$ mJy, the uncertainty is obtained adding in quadrature 
the measurements dispersion and the average of the individual errors.
The above flux was used to estimate the radio luminosity.

\subsection{HD\,145501}
\label{radio_145501}
Multifrequency radio measurements ($\nu=1.4$, 5, and 8.4 GHz) of HD\,145501 were reported by \citet*{linsky_etal92}.
The detection at 5 GHz was confirmed by \citet*{leone_etal94}.
This star has been also detected at 3 GHz ($S=1.45\pm0.2$ mJy) by the VLASS (VLASS1.2 tile T06t25).
The measured fluxes levels are almost comparable within the explored frequency range,
the corresponding average flux is $1.6 \pm 0.3$ mJy, which was used to estimate the luminosity of HD\,145501 at the
mean frequency $<\nu>\approx 3.6$ GHz.

\subsection{HD\,147932}
\label{radio_147932}
\begin{table}
\caption[ ]{HD\,147932 radio spectrum.}
\label{tab_147932}
\footnotesize
\begin{tabular}{c c l}
\hline
$\nu$                 &$<S_{\nu}>$                          &               \\
(GHz)                &(mJy)                        &notes                          \\
\hline       
$1.6$               &$9\pm4.5$            &  this paper   \\
$2.6$               &$15\pm4$            &  this paper   \\
$5.5$                  &$20\pm4$              & this paper                     \\  
$9$               &$21\pm4$              & this paper       \\  
$16.7^{\ast}$         &$20.7\pm0.2$              & \citet{leto_etal20b}             \\  
\hline
\end{tabular}
\begin{list}{}{}
\item[] $^{\ast}$ Flux from the total map (see \citealp{leto_etal20b}). 
\end{list}
\end{table}
Among the hot magnetic stars already known as sources of coherent auroral radio emission (ARE),
at the present time, HD\,147932 shows the unique behavior to have the auroral radio emission always detectable.
This star does not evidence any preferred phase where  
the low frequency highly polarized coherent pulses were detected \citep{leto_etal20b}.
In fact, its peculiar geometry makes always visible the highly beamed maser emission responsible for the ARE.
To disentangle the components arising from the two emission mechanisms,
incoherent (gyro-synchrotron emission) and coherent (cyclotron maser), 
as a first approximation, we assume all the circularly polarized flux produced by the ARE.
Hence, the subtraction of the average polarized flux (ATCA
data available in \citealp{leto_etal20b}) from the average total intensity 
supplies a reliable estimation of the incoherent radio spectrum of HD\,147932.
The fluxes, estimated at each observing frequency with the corresponding error propagation,  
are listed in Table~\ref{tab_147932}.
HD\,147932 was also observed by the VLA, 
source detected at 1.4 GHz by the NVSS ($S=9.4\pm0.6$ mJy; \citealp{condon_etal98}) 
and by the project AK460 (NRAO image archive, $S\approx 18$ mJy), 
and at 3 GHz, VLASS1.2 (tile T05t25) $S\approx 12$ mJy, 
but the fraction of their polarized emission is unavailable.
Therefore, considering the peculiar behavior of HD\,147932, these VLA data have not been used. 
The radio spectrum of the gyro-synchrotron emission component is shown in Fig.~\ref{spe10sou},
the corresponding average flux has been used to estimate the stellar radio luminosity at the
mean frequency $<\nu>\approx 5.1$ GHz.

\subsection{HD\,147933}
\label{radio_147933}
\begin{table}
\caption[ ]{HD\,147933 radio spectrum.}
\label{tab_147933}
\footnotesize
\begin{tabular}{c c l}
\hline
$\nu$                 &$<S_{\nu}>$                          &               \\
(GHz)                &(mJy)                        &notes                          \\
\hline       
$1.6$               &$5.5\pm2$            &  this paper; NVSS;  AK460$^{\dag}$  \\
$2.6$               &$7\pm2$            &   this paper;  VLASS1.2 (tile T05t25)    \\
$5.5$                  &$7\pm2$              & \citet{leto_etal20}                     \\  
$9$               &$8\pm2$              & \citet{leto_etal20}       \\  
$16.7$         &$8\pm2$              & \citet{leto_etal20}             \\  
$21.2$         &$7.5\pm2$              & \citet{leto_etal20}             \\  
\hline
\end{tabular}
\begin{list}{}{}
\item[] $^{\dag}$ NRAO image archive.
\end{list}
\end{table}
The behavior at the radio regime of HD\,147933 has been recently studied using 
multi-epoch and multifrequency ATCA observations \citep{leto_etal20}.
This star show coherent low frequency pulses superimposed to the basal incoherent non-thermal emission.
The ATCA observations were performed using wide band backends (bandpass 2 GHz wide at each observing frequency).
To better analyze the low frequency spectral dependence of the stellar radio emission, in this paper,
we reanalyzed the L-band ATCA measurements of HD\,147933.
This dataset has been split in two sub-bands, 1 GHz wide, centered at 1.6 and 2.6 GHz.
The low frequency pulses were observed at narrow phases intervals,
which were excluded from the data average.
The radio emission of HD\,147933 was also detected by the VLA
at 1.4 GHz, by the NVSS ($S=10.8 \pm 0.6$ mJy; \citealp{condon_etal98}) and unpublished archival data (project AK460: $S=8.9 \pm 0.3$ mJy),
and at 3 GHz (VLASS: $S=7.7\pm0.2$ mJy).
The VLA measurements have been averaged with the ATCA data at the closest frequency.
The average radio spectrum of HD\,147933 is listed in Table~\ref{tab_147933} and pictured in Fig.~\ref{spe10sou}.
The average of these spectral measurements has been used to estimate the radio luminosity reported in Table~\ref{tab_stars} 
at the mean frequency $<\nu>\approx 6.5$ GHz.

\subsection{HD\,170000}
\label{radio_170000}
The radio emission of this magnetic star was detected at 8.4 GHz ($S=0.45$ mJy; \citealp*{drake_etal06}).
HD\,170000 has been also detected by the VLASS at 3 GHz (VLASS1.1 and 2.1, tile=T28t10; fluxes respectively: $S=0.5\pm0.1$ and $S=0.8 \pm 0.1$  mJy).
These radio measurements have been averaged to estimate the radio luminosity of HD\,170000.
Mean frequency
$<\nu>\approx 5$ GHz.

\subsection{HD\,175362}
\label{radio_175362}
This star was detected at 5 GHz by \citet*{linsky_etal92} in two distinct epochs.
The average flux ($S=0.32 \pm 0.1$ mJy,
uncertainty estimated adding in quadrature the data dispersion and the average map noise)
is used to calculate the radio luminosity of HD\,175362.

\subsection{HD\,176582}
\label{radio_176582}
HD\,175582 was detected at 8.4 GHz ($S=0.46$; \citealp*{drake_etal06}).
Further, stellar radio emission was tentatively detected at 3 GHz
(VLASS1.1, tile=T20t24, $S\approx 0.3$ mJy).
To estimate the radio luminosity of this star we used the average flux $S = 0.4 \pm 0.1$ mJy, 
at the mean frequency $<\nu>\approx 5$ GHz.

\subsection{HD\,182180}
\label{radio_182180}
\begin{table}
\caption[ ]{HD\,182180 radio spectrum.}
\label{tab_182180}
\footnotesize
\begin{tabular}{c c l}
\hline
$\nu$                 &$<S_{\nu}>$                          &               \\
(GHz)                &(mJy)                        &notes                          \\
\hline       
$1.4^{\ast}$    &$7.9\pm0.6$            &  \citet{condon_etal98}   \\
$3^{\ast}$        &$11.3\pm0.2$             & VLASS1.1 (tile T04t30)              \\
$6$               &$16\pm3$             & \citet{leto_etal17}                      \\  
$10$               &$17\pm4$             & \citet{leto_etal17}                      \\  
$15$                &$16\pm3$              & \citet{leto_etal17}                       \\  
$22$                &$16\pm3$                   & \citet{leto_etal17}                      \\  
$33$                &$15\pm3$                   & \citet{leto_etal17}                      \\  
$44$                    &$13\pm2$              & \citet{leto_etal17}              \\  
\hline
\end{tabular}
\begin{list}{}{}
\item[] $^{\ast}$ Less than three measurements.
\end{list}
\end{table}
This is a bright radio source showing a clear rotational modulation \citep{leto_etal17}.
The stellar radio light curves have been performed in the range 6--44 GHz.
The average fluxes at each observing frequency are listed in Table~\ref{tab_182180}.
At lower frequencies HD\,182180 has been detected by the VLA all sky surveys
at 1.4 GHz (NVSS) and at 3 GHz (VLASS), the measured fluxes are also reported.
The average spectrum is shown in Fig.~\ref{spe10sou} and
the average flux has been used to estimate the corresponding radio luminosity.
Mean frequency
$<\nu>\approx 10.2$ GHz.

\subsection{HD\,215441}
\label{radio_215441}
This is the Babcock's star, that has a very high mean magnetic field strength ($\approx 34$ kG; \citealp{babcock60}).
Measurements of the HD\,215441 radio emission
were firstly reported by \citet*{linsky_etal92}, which observed at 1.4, 5, and 15 GHz using the VLA.
Another detection at 5 GHz was reported by \citet*{leone_etal94}.
The low frequency side of the stellar spectrum was explored by \citet{chandra_etal15}.
The authors reported repeated GMRT measurements,  at 0.61 and $\approx1.4$ GHz, evidencing variability
ascribed to the stellar rotation.
The average flux at $\nu=0.61$ GHz is $S=0.98\pm0.1$ mJy, 
at $\nu=1.4$ GHz is $S=1.3\pm0.2$ mJy \citep*{chandra_etal15,linsky_etal92}.
A similar flux level was detected at 3 GHz ($S\approx1$ mJy) by the VLASS1.2, tile=T24t23.
At higher frequencies the few existing measurements suggest a flux decreasing
($S\approx 0.85$ mJy, average of the two available data at 5 GHz, and $S\approx 0.5$ mJy at 15 GHz).
Using all the available radio measurements, we estimate the average flux of $S=0.9 \pm 0.3$ mJy,
the uncertainty has been estimated adding in quadrature the data dispersion and the mean value of the individual errors.
The above flux value has been used to estimate the radio luminosity of the Babcock's star at the
mean frequency $<\nu>\approx 2.9$ GHz.

\section{Radio emission of the selected Ultra Cool Dwarfs}
\label{ucd_appendix}

\subsection{BRI\,0021-02}
\label{radio_bri0021}
The incoherent radio emission from BRI\,0021-02 has been detected at $\nu=8.46$ GHz
by \citet{berger02}. 
The average incoherent basal flux is $S = 0.08 \pm 0.02$ mJy 
and was used to estimate the radio luminosity reported in Table~\ref{tab_ucd}.

\subsection{LSPM\,J0036+1821}
\label{radio_2m0036}
Highly polarized enhanced emission at 8.4 GHz was reported by \citet{berger02}
and by \citet{hallinan_etal08},
that is explained by the ECM emission mechanism.
Persistent radio emission due to the incoherent gyro-synchrotron emission
was also reported \citep{berger_etal05}.
The average incoherent flux measured at $\nu=4.86$ GHz 
is $S = 0.26 \pm 0.02$ mJy, value reported by \citet{metodieva_etal17}, 
with the corresponding references of the non-flaring emission.
The average flux was adopted to estimate the stellar radio luminosity of the incoherent emission.

\subsection{DENIS\,J1048.0-3956}
\label{radio_denis1048}
A very strong fully polarized radio pulse from DENIS\,J1048.0-3956 was observed at 8.64 GHz by \citet{burgasser_putman05},
this event was explained as clear signature of the coherent ECM emission. 
This radio pulse is superimposed above a basal quiescent incoherent non-thermal emission.
\citet{ravi_etal11} collected multifrequency radio observation of DENIS\,J1048.0-3956
performing the spectral distribution of the stellar incoherent non-thermal radio emission up to 24 GHz.
The spectrum has a power law shape with a negative spectral index ($\approx -1.71$).
Reliable estimation of the average flux from DENIS\,J1048.0-3956 within a spectral range 5.5--10 GHz
is $\approx 0.2$ mJy with uncertainty of 50\%.
The above flux value was used to estimate the radio luminosity of DENIS\,J1048.0-3956
reported in Table~\ref{tab_ucd}.

\subsection{TVLM\,513-46}
\label{radio_tvlm513}
Periodic fully polarized coherent pulses from TVLM\,513-46, due to the ECM emission, 
was clearly observed at 8.44 and 4.88 GHz by \citet{hallinan_etal07}.
It was also suggested that the cyclic behavior of the multi-frequency coherent pulses from this dwarf star 
is a possible sign of maser emission triggered by an exo-planet in close orbit \citep{leto_etal17b}. 
Besides the ECM coherent emission,
also the incoherent radio emission component of TVLM\,513-46 exists.
The radio spectrum of the incoherent emission from TVLM\,513-46
has been studied up to the high frequency spectral range covered by ALMA 
\citep{williams_etal15b}.
The average flux over the spectral range $\approx1.6$--90 GHz
\citep{osten_etal06,williams_etal15b}, mean frequency
$<\nu>\approx 7.2$ GHz, is $ S \approx 0.2$ mJy and was adopted to estimate the radio luminosity of 
the incoherent radio emission of TVLM\,513-46.

\subsection{LSR\,J1835+3259}
\label{radio_lsr1835}
This M8.5 dwarf showed periodic fully polarized pulses due to the coherent ECM emission mechanism \citep{hallinan_etal08}.
These are likely modulated by an orbiting exoplanet \citep{hallinan_etal15}. The coherent pulses of LSR\,J1835+3259
are superimposed above a basal persistent incoherent gyro-synchrotron emission \citep{hallinan_etal08,berger_etal08}.
The average incoherent flux of $S = 0.525 \pm 0.015$ mJy, detected several times at $\nu=8.46$ GHz,
has been reported by \citet{metodieva_etal17}, 
with the corresponding references.
The above average flux was used to estimate the corresponding radio luminosity of LSR\,J1835+3259.

\subsection{V374\,Peg}
\label{radio_v374peg}
In three observing epochs of January 2007, the VLA observed V374\,Peg at $\nu=8.4$ GHz \citep{hallinan_etal09}.
These observations phased with the stellar rotation period allowed to obtain
the radio light curve of V374\,Peg \citep{llama_etal18}. The stellar radio emission
showed a clear rotational modulation.
The measured flux varies between $\approx 0.4$ and $\approx 1$ mJy.
To estimate the radio luminosity of V374\,Peg (reported in Table~\ref{tab_ucd}) 
we roughly used the average flux $S\approx 0.7$ mJy.

\section{The equations of the wind}
\label{wind_appendix}

{
For a fully ionized gas (hydrogen 90\%, helium 10\%, corresponding to a mean ionic charge $Z \approx 1.1$)
the expected radio emission from the stellar ionized wind can be derived following the relation \citep{scuderi_etal98}: 

\begin{equation}
\label{eq_wind_spectrum}
S_{\nu}=7.26 \times
\left(\frac{\nu}{{10\,\mathrm{GHz}}}\right)^{0.6}  \times
\left(\frac{0.85 \,T_{\mathrm{eff}}}{10^4 \,{\mathrm K}}\right)^{0.1} \times
\end{equation}
\begin{displaymath}
\left(\frac{\dot M}{10^{-6} \,{\mathrm {M_{\odot} \, yr^{-1}}}}\right)^{4/3} \times
\left(\frac{1.3 \, v_{\infty}}{100 \,{\mathrm {km \, s^{-1}}}}\right)^{-4/3} \times
\left(\frac{D}{{10^3\,\mathrm{pc}}}\right)^{-2}  
\mathrm{~~(mJy)}
\end{displaymath}

Following the above reported assumption for the ionized gas composition,
the equation given by \citet{panagia_felli75} for the effective radius  of a spherical wind emitting region
is:

\begin{equation}
\label{eq_wind_radius}
R_{\nu}=10434 \times
\left(\frac{\nu}{{10\,\mathrm{GHz}}}\right)^{-0.7}  \times
\left(\frac{0.85 \,T_{\mathrm{eff}}}{10^4 \,{\mathrm K}}\right)^{-0.45} \times
\end{equation}
\begin{displaymath}
\left(\frac{\dot M}{10^{-6} \,{\mathrm {M_{\odot} \, yr^{-1}}}}\right)^{2/3} \times
\left(\frac{1.3 \, v_{\infty}}{100 \,{\mathrm {km \, s^{-1}}}}\right)^{-2/3}
\mathrm{~~(R_{\odot})}
\end{displaymath}

}

\end{document}